\documentclass[structabstract]{aa} 
\usepackage{graphicx}
\usepackage{txfonts}
\usepackage{natbib}
\begin{document}
   \title{The large-scale environment of low surface brightness galaxies}


    \author{S.\,D. Rosenbaum\inst{1,2}
	  \and
            E. Krusch\inst{3,2}
          \and
          D.\,J. Bomans\inst{2}  
           \and
          R.-J. Dettmar\inst{2} 
          }

\offprints{S.\,D. Rosenbaum}

   \institute{German Aerospace Center (DLR), Remote Sensing Technology 
	      Institute, Photogrammetry and Image Analysis Unit, M\"unchner Str. 20, 82234 We{\ss}ling, Germany\\\email{dominik.rosenbaum@dlr.de}
	      \and
	      Astronomisches Institut, Ruhr-Universit\"at Bochum (AIRUB),
              Universit\"atsstr. 150, 44780 Bochum, Germany
              \and
	      German Aerospace Center (DLR), German Space Operation Center (GSOC), Mission Operations Division,  M\"unchner Str. 20, 82234 We{\ss}ling, Germany}


   \date{Received March 13, 2007; accepted July 2, 2009}

  \abstract{The exact formation scenarios and evolutionary processes that led 
to the existence of the class of low surface brightness galaxies (LSBs) have  
not yet been understood completely. 
There is evidence that the lack of star formation expected to be 
typical of LSBs can only occur if the LSBs were formed in low-density 
regions.}
{Since the environment of LSBs has been studied before only on small scales (below
2\,Mpc), a study of the galaxy content in the vicinity of LSB galaxies on 
larger scales  could add a lot to our understanding of 
the origin of this galaxy class.}
{We used the spectroscopic main galaxy sample of the SDSS DR4
to investigate the environmental galaxy density of LSB galaxies 
compared to the galaxy density in the vicinity of high surface
brightness galaxies (HSBs). 
To avoid the influence of evolutionary effects depending of redshift
and to minimize completeness issues within the SDSS,  
we limited the environment studies to the local universe with a redshift of 
$z\leq0.1$. At first we studied the luminosity distribution
of the LSB sample obtained from the SDSS within two symmetric redshift intervals
($0.01<z\leq0.055$ and  $0.055<z\leq0.1$).} 
{It was found that the 
lower redshift interval is dominated by small, low-luminosity LSBs, whereas
the LSB sample in the higher redshift range mainly consists of
larger, more luminous LSBs. This comes from selection effects of the SDSS
spectroscopic sample.
The environment studies, also divided into these two redshift bins, 
show that both the low mass, and the more massive
LSBs possess an environment with a lower galaxy density than HSBs.
The differences in the galaxy density between LSBs and HSBs 
are significant on scales between 2 and 5\,Mpc, the scales of groups
and filaments.
To quantify this, we have introduced for the first time 
the LSB-HSB Antibias. The obtained LSB-HSB Antibias parameter 
has a value of $10\%$-$15\%$.} 
{From these results we conclude that LSBs formed in low-density
regions of the initial universe and have drifted until now 
to the outer parts of the filaments and walls of the large-scale structure.
Furthermore, our results, together with actual cosmological simulations,
show that LSBs are caused by a mixture of nature and nurture.}


   \maketitle%

\section{Introduction}
Although it is clear today that low surface brightness galaxies (LSBs) 
are a more common class of galaxies (e.g. \citealt{sprayberry94, de_jong96,
o_neil00b}) than expected in the beginnings of LSB 
research (e.g. \citealt{freeman70}), their formation and evolution scenarios 
have so far not been understood well (e.g. \citealt{impey97}).
Three decades ago, the appearance of the LSB phenomenon was thought to be
rare, too rare to significantly contribute to the total galaxy content of the
universe.  
\cite{freeman70} studied the central surface-brightness distribution
in Johnson $B$ band of 36 spiral and S0 galaxies. He found this distribution 
to be
fitted well with a Gaussian distribution with a peak at 21.65\,mag/arcsec$^2$ 
and a width of $\sigma=0.3$\,mag/arcsec$^2$.
\cite{disney76} turned the attention to selection effects of the
galaxy surveys. He used the selection effects to
explain this sharp surface-brightness distribution found by  \cite{freeman70}.
He also mentioned that our picture of the universe is biased towards
galaxies with high surface brightnesses due to those selection effects.
\cite{fisher81} published
H\,I observations of a large galaxy sample and compared their
detections to the Catalogue of Galaxies and Clusters of Galaxies (CGCG, 
\citealt{zwicky68}). They found that numerous LSBs fell below the magnitude
cutoff in the CGCG. Later it was realized (e.g. \citealt{romanishin82})
and became more popular that 
a distinct class of LSB galaxies with different properties exists apart 
from the class of the normal HSB galaxies, which were thought to 
form the dominant galaxy population of the universe.  
A lot of work on the 
number density of the faint-end surface-brightness distribution 
has been done (e.g. by: \citealt{mcgaugh95}, \citealt{dalcanton97},
\citealt{o_neil00b},\citealt{o_neil03}) to show that the number 
density 
of LSB galaxies is not only more than a $3\sigma$ phenomenon, but also
comparable to the number density of galaxies at high surface brightness.

Since the introduction of a separation between LSBs and HSBs 
was in the beginning a working hypothesis to show that
the \cite{freeman70} law is wrong at the faint end, 
the definition of low surface brightness is not used consistently
in the literature. It varies from $\mu_{B}(0)$=22.0\,mag/arcsec$^2$ 
(e.g. \citealt{mcgaugh95}, i.e. 1.33$\sigma$ beyond the Freeman value)
 up to $\mu_{B}(0)$=23.0\,mag/arcsec$^2$ (as used for instance in  
\citealt{impey97}, i.e. 4.66$\sigma$ beyond the Freeman value).

In the present publication, a balance between these two extremas
has been set, and a value of  $\mu_{B}(0)$=22.5\,mag/arcsec$^2$ 
was chosen to be the border between low and high surface brightness.
This definition has already been used by other authors before
(e.g. \citealt{de_blok95, morshidi99, meusinger99, rosenbaum04}).

It is known that giant LSBs possess gas masses similar to those of high
surface brightness galaxies (HSBs) (e.g. \citealt{pickering97}
\citealt{matthews01, o_neil02}, 
and \citealt{o_neil04}).
However, the gas mass converted into stars must be lower in LSBs than in HSBs.
This results in a higher gas mass fraction for LSBs than for 
HSBs. \cite{mcgaugh97} examined the gas mass fraction of spiral galaxies
in the context of surface brightness. 
The gas mass fraction  is defined as $f_g=M_g/(M_g+M_*)$ (with 
$M_g$ the total gas mass and $M_*$ the stellar mass) and describes the
fraction of gas mass that has not (yet) been converted into
stars. The authors found  this fraction
to be significantly higher in LSBs than in HSBs. 
These facts all contribute to a picture of LSBs being equipped with enough 
gas to form as much stars as HSBs, but obviously it did not happen. 

This may be connected with the low gas surface density found in LSBs.
As found by e.g. \cite{van_der_hulst93} or \cite{pickering97}, 
the surface density of the large
LSBs of their sample is systematically below the critical density for the
formation of molecular, star-forming clouds (\citealt{kennicutt89}, known
as Kennicutt criterion). 
Thus, the key to the understanding of
these LSB galaxies rests in the answer to the question what kept the 
surface density to stay below this critical value. One clue to this question 
comes from studies of the environment. If LSBs have no galaxy 
neighbors on small and intermediate scales, this absence of a gravitational
trigger will be able to keep the gas in a stationary situation.
This mode would be without much 
turbulence and hence without density perturbations which cause gas clouds
to collapse and to initiate sufficient star formation.

For this evolutionary scenario, evidence already exists in the literature.
A lack of nearby neighbors
on small scales below 2\,Mpc has already been found by \cite{bothun93}, 
who examined the
spatial distribution of LSB galaxies in the Center for Astrophysics redshift 
survey (CfA, \citealt{huchra93}). They  
performed galaxy number counts within cones of a mean projected
radius of 0.5\,Mpc and a velocity range of 500\,kms$^{-1}$. 
These results were validated by \cite{mo94} who studied the spatial
distribution of LSBs by calculating the cross correlation functions of LSBs
with HSBs of the CfA and IRAS sample \citep{rowan91}. The fact that nearby 
($r\leq0.5$\,Mpc) companions of LSB galaxies are missing was also detected by 
\cite{zaritsky93}. The idea that a lack of tidal interaction in LSB disks
causes a suppression of star formation and keeps the system unevolved,
fits also to the results of theoretical models of tidally triggered galaxy 
formation (e.g. \citealt{lacey91}).
Other clues  to the theory of LSB galaxies 
forming in a less dense
environment than HSB galaxies come from color and metallicity
studies. Most LSBs are found to be blue. Their $(B-V)$ color index
shows an average value of $(B-V)\simeq0.44$\,mag \citep{mcgaugh94b, 
romanishin83}.
This is an effect of the mean age of the dominant stellar population 
\citep{haberzettl06}. It cannot be an effect of metallicity,
since there is no correlation between the oxygen abundance and the
color of the galaxies \citep{mcgaugh94b}. Moreover, \cite{pickering95}
found most of the Malin-like very massive LSBs of a sample of 10 objects to 
possess
metallicities of  $Z=Z_\odot$. The low probability of external triggers 
for star formation in LSBs located in low-density environments would
result in a star formation initiating at a later date compared to that
of HSBs, if the low-density scenario is true. This would then result in
a younger dominant stellar population in LSBs and bluer colors
(as observed by \citealt{haberzettl06}) since in this LSB formation 
scenario galaxy formation first took place in the overdense regions of the 
initial universe and later in areas with lower densities.

Another plausible explanation for the existence of LSBs comes from 
dark matter simulations of disk galaxy formation scenarios (e.g. \citealt
{dalcanton97b, boissier03}). In these simulations, the dark 
matter halos of LSB disks were found to have a higher spin parameter than that 
of HSB spirals.
The spin parameter $\lambda$ is a dimensionless quantity. 
It describes how much angular momentum of the dark matter halo 
is transferred to the disk. 
In the study by \cite{boissier03} the spin parameter of LSBs with 
$\mu_B(0)>22.0$\,mag/arcsec$^2$ exceeded values of $\lambda=0.06$.
The higher spin parameter of the LSB Dark Matter halos implies that more
angular momentum is contained in the disk which could naturally result in 
a larger ratio of scale lengths to luminosity in LSB disks. 
This means that the total amount of 
gas (which is similar in mass to that of HSBs) is distributed on larger
scale lengths than in HSBs. This scenario would explain the low gas surface 
densities, which cause the galaxies to be LSBs due to the nature of their dark 
matter halos and stays in competition with the scenario that LSBs formed
in low-density environments.
The discussion which of these two distinct possible explanations for the
existence of LSBs is real is often reduced to the question: 
Are LSBs caused by nature or nurture? The aim of the present 
work is to  throw some light onto this question.


\section{Data Characteristics and Analysis}
\label{LSB_selection}
The Sloan Digital Sky Survey (SDSS, \citealt{york00}) is a large imaging and 
spectroscopic survey. It covers in one of the actual public data releases   
--namely DR4 \citep{dr4} which was used in the present work--   
around 6700\,deg$^2$ in imaging and 4783\,deg$^2$ in 
spectroscopy.  The survey area of the SDSS is mainly distributed in
two equatorial regions and one northern cap scan region. 
Data were taken
with a 2.5\,m telescope located at the Apache Point observatory, New Mexico.   
Two instruments are used at this telescope.
For the imaging mode a complex camera containing 30 charge-coupled devices 
(CCDs) 
is mounted onto the telescope.
With this camera it is possible to obtain images simultaneously in 
the modified Gunn-bandpasses $u,g,r,i,z$ (the photometric system is 
described in detail by \citealt{fukugita96}). 
The special time-delay-and-integrate operation mode 
and the architecture of the CCD camera in connection with the f/5 focus ratio
results in an exposure time of 54\,s for each filter.

For spectroscopy a pair of two channel fiber optics spectrographs, which are 
able to 
take spectra of 640 objects simultaneously, is operated at the survey 
telescope. 

The spectrographs provide a typical 
resolution of $\lambda / \Delta\lambda\simeq1800$, which translates to a
value of 167\,km/s in velocity resolution. The pixel size in velocity space 
is 69\,km/s.   
The fiber diameter 
corresponds to an angular size of 3''. 
Spectroscopy is undertaken with guided exposures of overlapping plates.
Each plate has a projected diameter of 3\degr. 
All spectroscopic fields are obtained with a total integration time of  
45~minutes and more.
This results in a signal-to-noise
ratio of $(S/N)^2$=4.5\,pixel$^{-1}$ for an object with a magnitude of 20.2 in 
the $g$ band. 

\noindent A series of interlocking pipeline tasks processes the data 
automatically.  
Thereby, a special algorithm searches for
galaxies, quasars and stars and selects them for spectroscopic follow-up 
observations. 


For spectroscopic galaxy target selection, apparent
\cite{petrosian76} magnitudes in $r$-band are used as selection criterion. 
Petrosian magnitudes are a good measure for the total magnitude 
of galaxies, since they are 
independent from sky brightness, foreground extinction, the galaxy central 
surface brightness and Tolman dimming \citep{strauss02}.
The selection algorithm for galaxies chooses objects with
an $r$-Petrosian magnitude brighter than $r=17.77$\,mag. 
Additionally, an $r$-band Petrosian half
light surface brightness of $\mu_{50}\leq24.5$\,mag/arcsec$^2$ is 
required. These cut criteria select about 90 galaxies per square degree for
follow-up spectroscopy. The fraction of galaxies which were detected
in the SDSS images but are not chosen as a spectroscopic target 
due to the surface brightness limit is very small (0.1\%,
\citealt{strauss02}). 
Further the authors state, that the selection rules for total magnitude and 
surface brightness result in a completeness for galaxy targets of $\sim99\%$.
They also state that the loss of galaxies
for spectroscopic follow up due to fiber positioning constraints is
with ~6\% small, too. In general, galaxies are not rejected because of this 
constraint, but sometimes, preferably  in
dense environments like clusters this constraint applies.
This means in a total that $\sim93\%$ of the imaged galaxies are
selected for spectroscopic follow-up observations with the SDSS telescope. 
 
If the field is not very crowded and less than 
640 objects are selected by applying the strict 
selection rules for galaxies and stars, these criteria are softened.
Thus, also galaxies fainter than $r>17.77$\,mag or with a half light surface 
brightness fainter than $\mu_{50}>24.5$\,mag/arcsec$^2$
are selected.

The minimum distance for the placement of two adjacent fibers is 55'' in 
projection at the sky. 
If there are two or more possible candidate objects for SDSS follow-up 
spectroscopy within a distance to each other of 55'', not necessarily the 
brightest object is selected for spectroscopy. 
In \cite{blanton05} it is stated that in this case, one member is chosen 
independently from its magnitude or surface brightness.
This is a very important for performing environment studies
in dependence of surface brightness of the objects using the spectroscopic
main galaxy sample of the SDSS. 
It means, that in the case of fiber 
spacing constraints concerning a LSB and a neighboring HSB (both
fulfilling the Petrosian-$r$ and the  half light surface 
brightness criterion), not necessarily
the HSB is chosen for spectroscopy. The LSB has the same chance to be
preferred. This means that the environment studies are not biased against LSBs 
in dense environments due to fiber spacing constraints.


All these properties of the selection function do not directly apply
biases against LSB galaxies. Firstly, this is due to the fact 
that if two (or more) adjacent target candidates are closer to each other 
than 55'', the selection is done randomly and not necessarily the
brightest object is selected. Secondly, as stated by \cite{strauss02}, in the 
absence of noise, the Petrosian
aperture is not affected by external effects like foreground extinction,
the Tolman dimming and sky brightness. Thus, identical galaxies seen
at two different (luminosity) distances have fluxes related exactly as the
inverse square of distance (in the absence of $K$-corrections). 
They further argue, that one can 
therefore determine the maximum distance at which a galaxy would enter a 
flux-limited sample without knowing the galaxy's surface brightness profile
(which would be needed for the equivalent calculation with e.g. isophotal 
magnitudes). Hence they conclude that two galaxies which have the same surface 
brightness profile shape but different central surface brightnesses have the
same fraction of their flux represented in the Petrosian magnitude, so there
is no bias against the selection of low surface brightness galaxies of
sufficiently bright Petrosian magnitude. 

Further details of the SDSS data aquisition and reduction are found in
the publications of the corresponding data release (EDR: \citealt{stoughton02},
DR1: \citealt{dr1}, DR2: \citealt{dr2} DR3:  \citealt{dr3}, and DR4: 
\citealt{dr4}).

\subsection{Obtaining the dataset from the SDSS DR4}
For the present publication, the DR4 is the main data source.
The LSB candidates and the HSB comparison sample were retrieved from the 
spectroscopic main galaxy sample by downloading a set of parameters
from the DR4 database using the SDSS Query Analyzer tool. The following 
parameters of each object, classified as a galaxy with spectroscopic data 
available were downloaded: Object identifier, right ascension, 
declination, an azimuthally averaged radial surface brightness profile in the 
$g$- and $r$-band, the redshift of the object with its error, and the apparent
Petrosian-g and r magnitudes. The sample was limited to a  redshift 
of $z\leq0.11$. 

This redshift margin was applied
to limit the analysis to the local universe and to exclude the 
influence of evolutional effects from the 
environment studies which appear with redshift.
To minimize the uncertainty of the redshift
a z-confidence greater than 90$\%$ was demanded. 
With these 
selection criteria, a dataset containing a total of 212080 galaxies
of both Low and High Surface Brightness type was obtained.


\subsection{The SDSS as a survey for LSBs}
Galaxy surveys
are generally biased towards galaxies with higher surface brightnesses due
to selection effects (e.g. \citealt{disney76}), so is the SDSS. 
In a nutshell, selection effects are caused by the sky brightness at the telescope
site including lunar phases, angular moon distances to the objects, 
atmospheric conditions like seeing, and transparency, the optical properties 
of the telescope, flat-fielding, detector properties 
(like quantum efficiency, dark current and readout
noise) and,  
the sky brightness produced by our own galaxy and its 
absorption and extinction. 

All these effects bias our survey results
towards a large number of HSB galaxies and only a
sparse number of galaxies with low surface brightness.   
Additionally to this bias, there is a dimming effect of the surface
brightness of the individual galaxy in dependence of redshift \citep{tolman34}.

When one searches a survey for LSBs which means dealing with 
galaxies at the edge of the detection limit, it is important to understand
how complete the obtained catalogue is and what kind of objects were missed.
In the case of the SDSS, one has to understand which properties the 
galaxies have, which were chosen by its selection effects.
It is also important 
to know how complete the surface-brightness distribution is at the faint end. 
Therefore, the selection function of the 
spectroscopic main galaxy sample of the SDSS DR4 has to be known.
Galaxies in the survey are found by 
special algorithms, which 
investigate the data by
searching for signals above a certain noise deviation threshold and then
applying diameter criteria or magnitude limits or both to the measured
surface brightness profiles or total magnitudes. 
For automated spectroscopic surveys like the SDSS or the two-degree-Field
Galaxy Redshift Survey  (2dFGRS, \citealt{colless01}),
there are routines which automatically find target galaxies within the 
survey images and then assign fibers to the chosen objects.

In the dataset obtained as described above, a number of 1978 LSBs was
found. LSBs were detected by fitting exponential surface brightness profiles 
(following \citealt{o_neil97a}) to 
the SDSS measured azimuthally averaged surface brightness profiles in $g$- and 
$r$-band of each galaxy in the downloaded dataset. Then, the obtained central 
surface brightness was converted into a Johnson-$B$ magnitude and 
it was corrected for \cite{tolman34} dimming depending on the
redshift of the individual galaxy.

If the obtained (cosmological dimming corrected) value of the central $B$ 
surface brightness fulfilled the LSB criterion
\begin{eqnarray}\label{mue_profile}
   \mu_{B}(r=0)\geq22.5\,{\rm mag/arcsec}^2,
\end{eqnarray}
the galaxy was flagged as LSB in the dataset, otherwise it remained as HSB
galaxy.

However, the fraction of LSBs found in the SDSS DR4 data
with respect to the HSB galaxies, is low ($\sim0.65\%$), but still 3 times
higher than predicted by the \cite{freeman70} Law.

Hence, the photometry is the limiting factor on the
completeness of the spectroscopic main galaxy sample. Although the 
drift scan exposure mode provides an excellent flat fielding of the 
photometric data which should be sensitive to low surface brightness 
objects, the exposure times of 54\,s with a 2.5\,m telescope are too
short to be well suited for hunting LSB galaxies. Nevertheless, 
the SDSS is currently the only available survey which covers a volume large 
enough to perform environment analyses on the large scale 
distribution of LSBs. 

\begin{figure*}[t]
\begin{center}


\includegraphics[width=8.5cm]{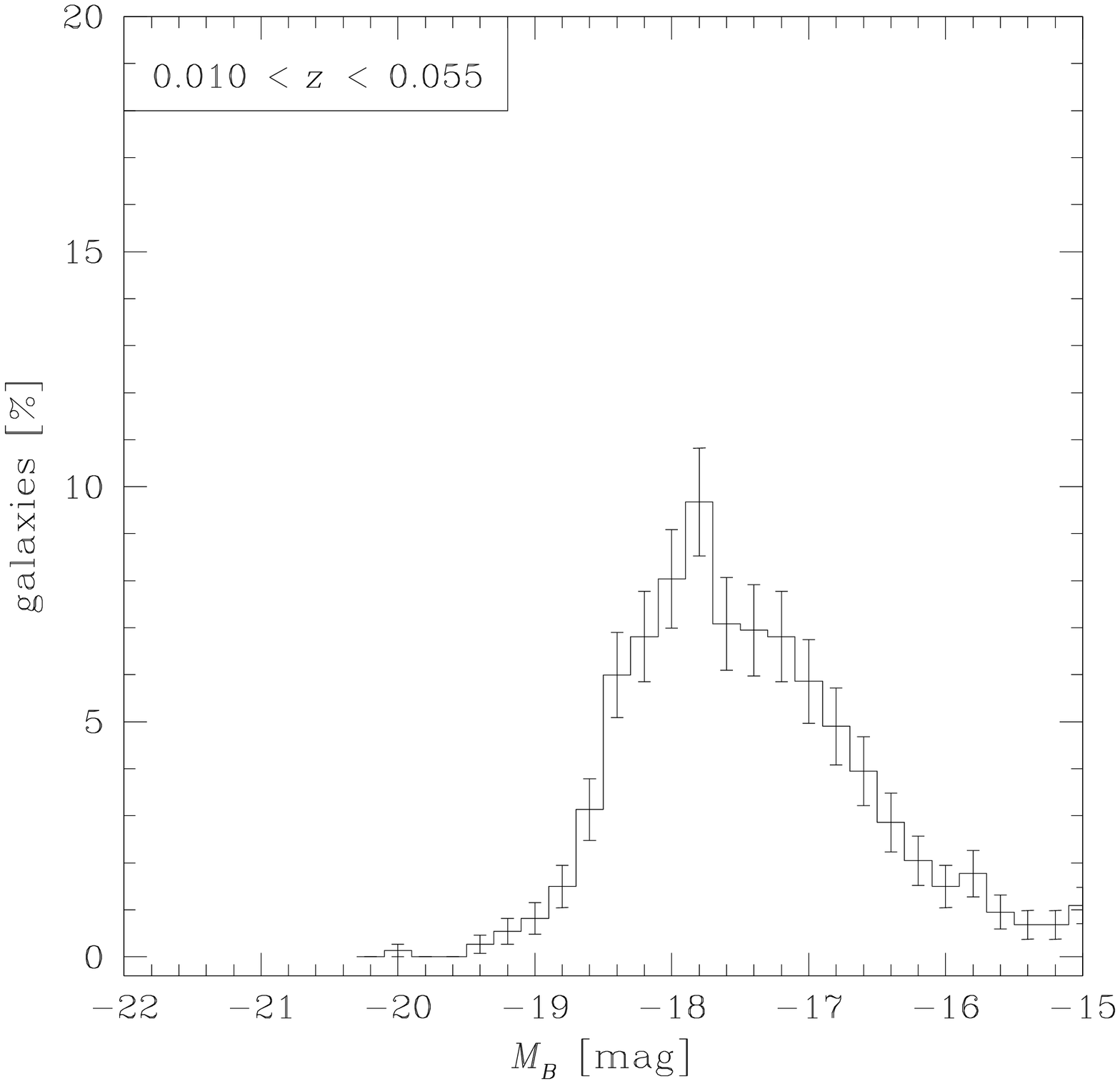}
\includegraphics[width=8.5cm]{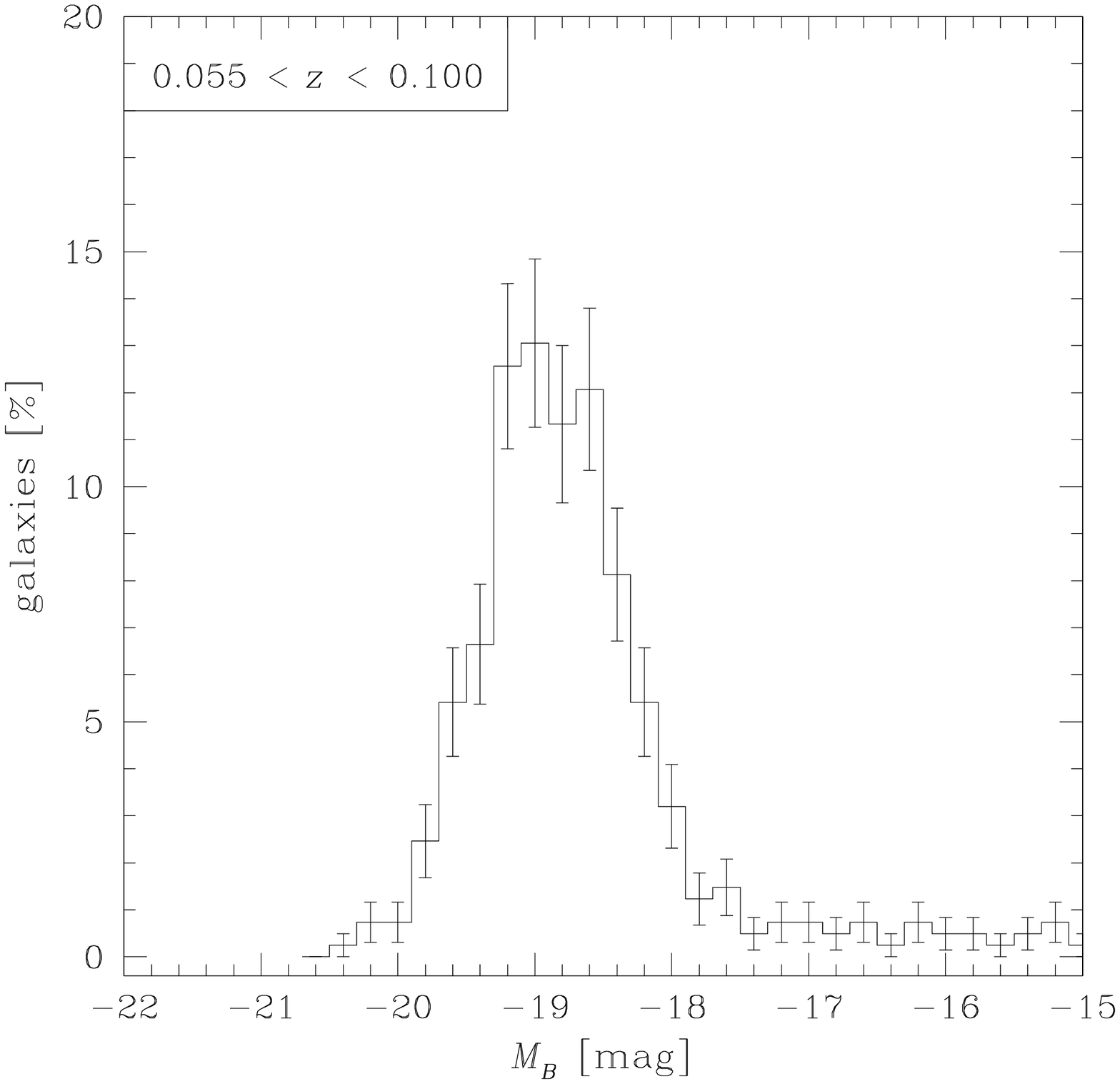}
\caption[LSB Luminosity Distribution]
{The distribution of the absolute magnitude of all sample LSB galaxies
with a redshift of $z\leq0.11$. Left panel shows the absolute
magnitude distribution of the LSB galaxies in a redshift interval of
$0.01<z<0.055$ in the filter Johnson $B$
(from top to bottom). It was calculated from the apparent Petrosian magnitudes
measured by the SDSS in $g$ and $r$ by applying a distance modulus obtained 
from the redshift and transforming it into Johnson B following \cite{smith02}. 
Right panel shows the according distribution 
for the $z$-range $0.055<z<0.1$. Mind that the peak of the 
distribution for the higher redshift interval is shifted towards in total
magnitude brighter (larger) galaxies in comparison to the lower redshift 
interval.}
\label{abs_mag_distrib}
\end{center}
\end{figure*}
  
Different studies find that
the total luminosity of the typical LSB contained in the sample 
varies with redshift caused by selection effects.
Furthermore, before dealing with the environment of LSB galaxies,
it is important to understand of what kind the LSBs obtained in the
resulting sample from SDSS DR4 are.
Therefore, the absolute Petrosian B magnitude was
calculated from the absolute Petrosian magnitudes in SDSS $g$ and $r$ bands
by using the transformation of \cite{smith02}. Before, the absolute $g$ and $r$
Petrosian magnitudes were calculated from the measured apparent quantities
by converting the redshift of each galaxy into a distance modulus. 
$K$ corrections (e.g. \citealt{humason56}, \citealt{oke68})
were not applied.
This was caused the fact that their influence on the absolute magnitude is in 
the order of magnitude of the photometric error for galaxies in the local 
universe and therewith negligible.  

Figure \ref{abs_mag_distrib} shows the histogram of the absolute magnitude 
distribution of all LSB sample galaxies, which were later used in the
environment studies, divided into two symmetrical redshift bins.
The left panel shows the relative number (in percent) versus the absolute 
magnitude in the filter Johnson $B$ within the
redshift interval of $0.01<z<0.055$. The right panel shows the 
absolute magnitude distribution for the same filter, but for the redshift 
interval of $0.055<z<0.1$.

\begin{figure*}[t]
\begin{center}
\includegraphics[width=7.5cm]{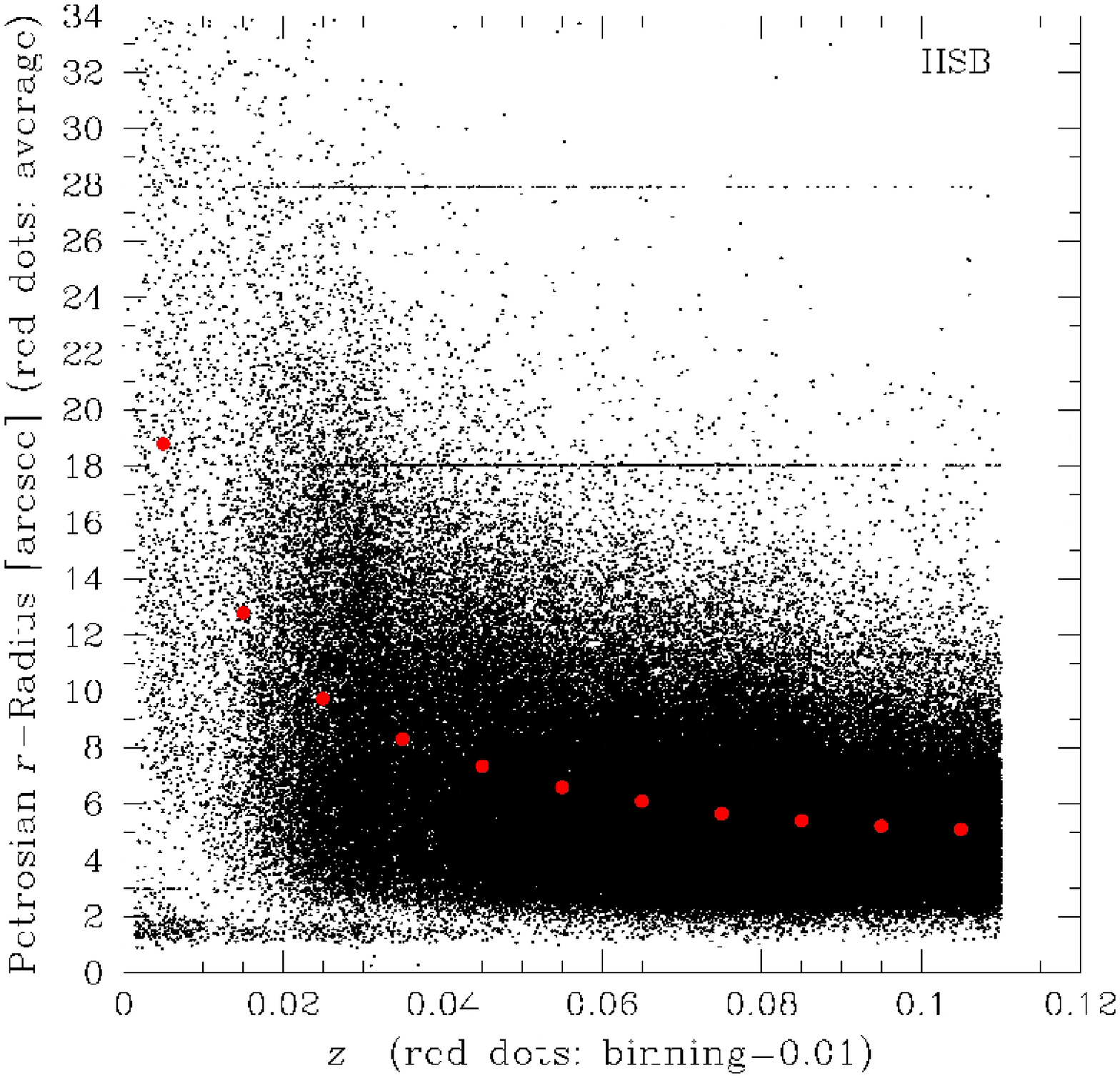}
\includegraphics[width=7.5cm]{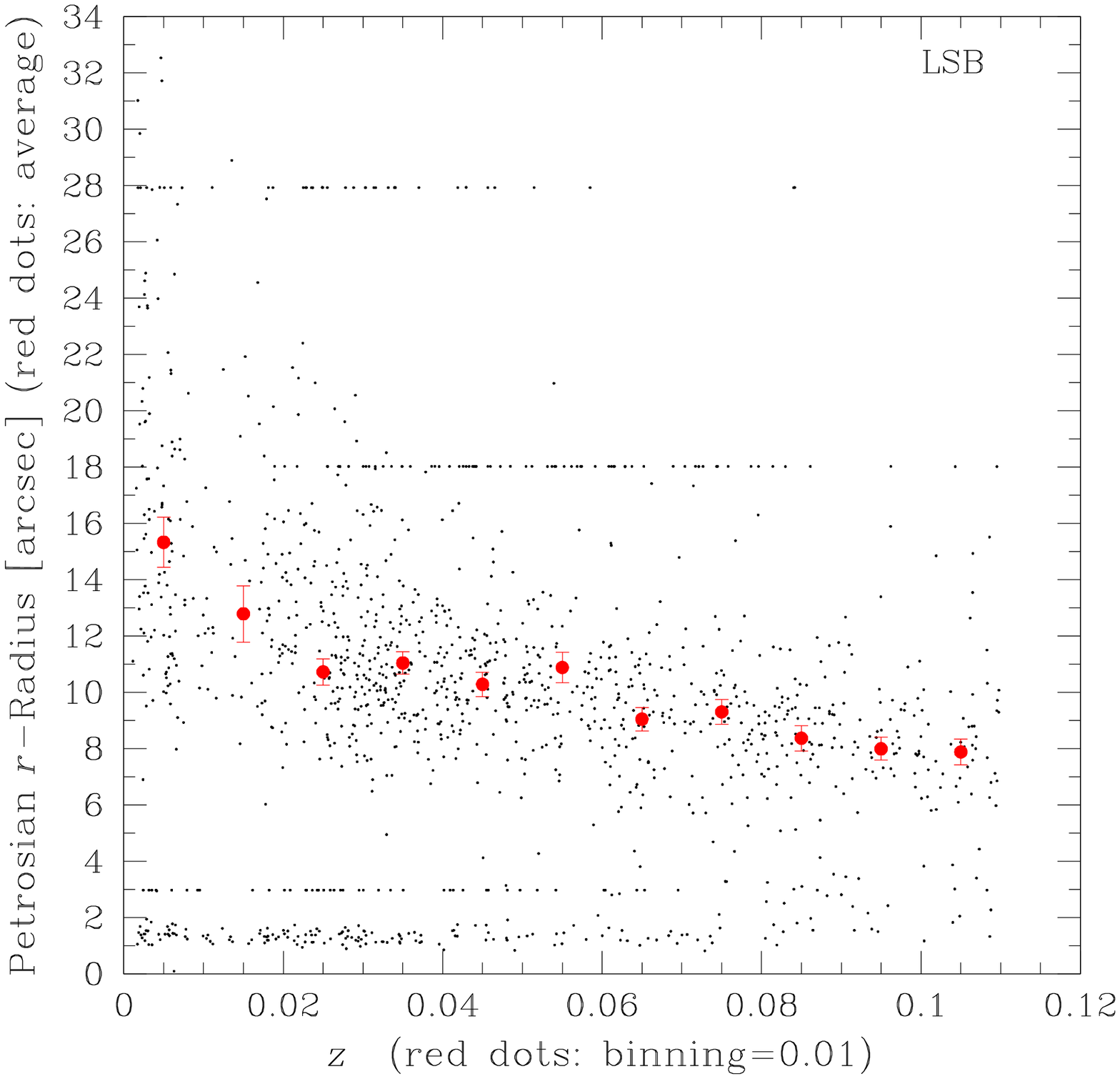}
\includegraphics[width=7.5cm]{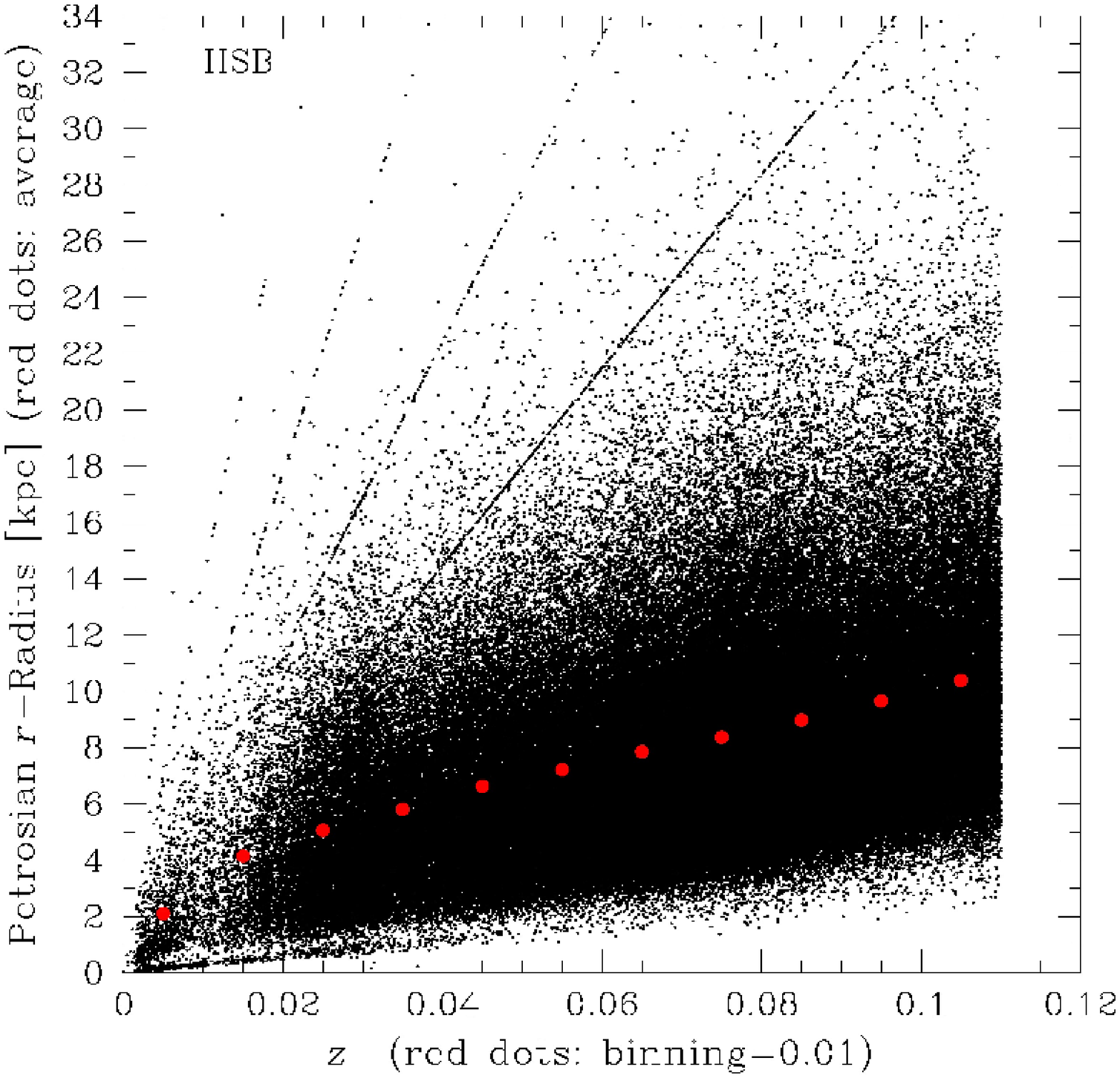}
\includegraphics[width=7.5cm]{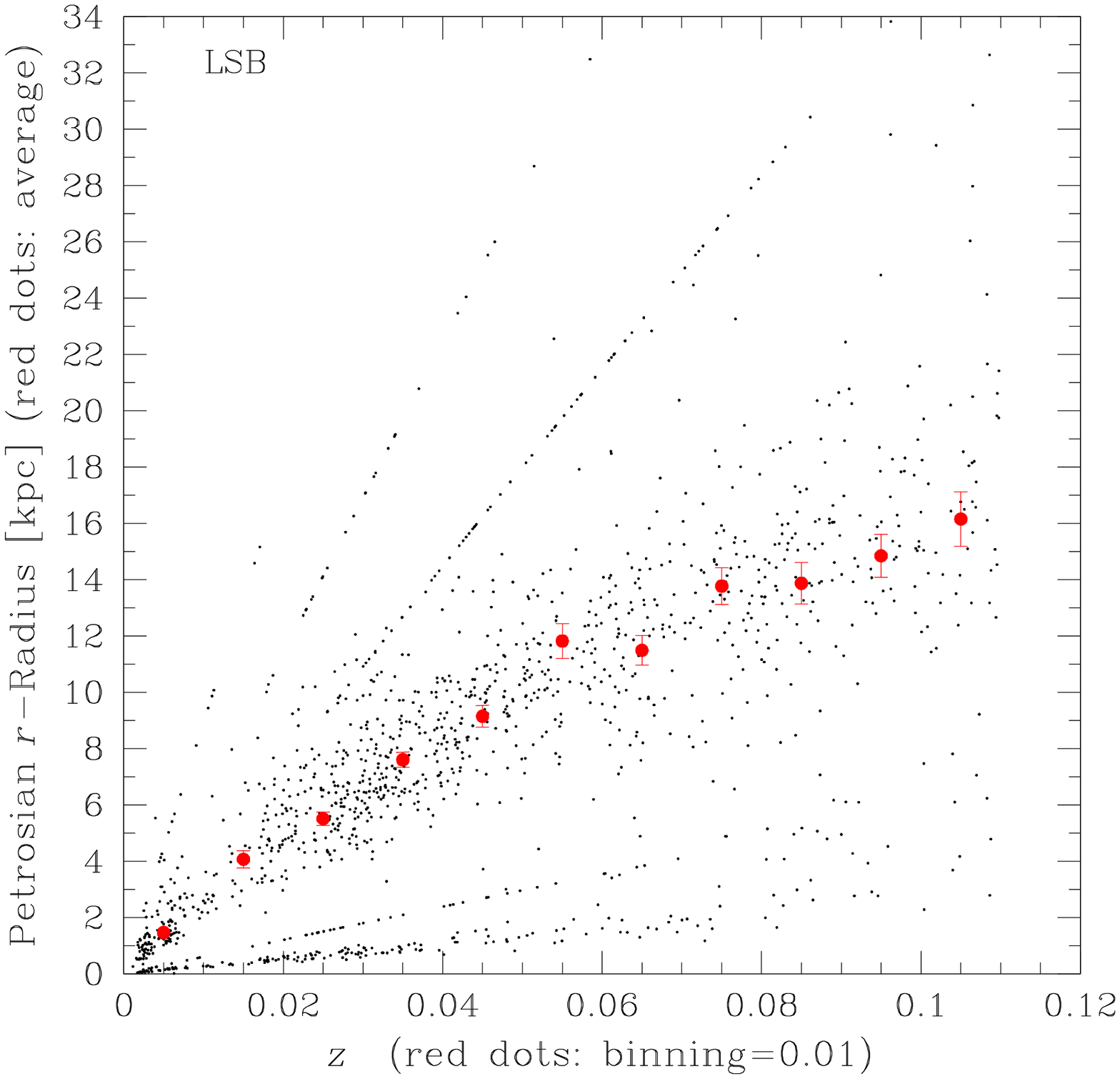}
\caption[Petrosian-$r$ Radius vs. Redshift of all Sample Galaxies]{
The left upper panel shows the distribution of the apparent Petrosian-$r$ 
radius
(in arcsec) against redshift $z$ of HSB sample galaxies (black dots). 
Red dots are the average
Petrosian radii (in arcsec) within redshift bins with a bin width of 0.01. 
The statistical error has a typical value of 0.07\,arcsec and is therewith 
smaller than the size of the red points.
Right upper panel shows the same as left upper panel, but for the LSB sample 
galaxies. Here, error bars indicate the statistical error of the
indiviual average.
Mind, that for a redshift of $z>0.01$ the LSB galaxies are larger
in apparent radius than the HSBs. The statistical error is smaller than the
size of the red points.

At the lower left panel the distribution of the absolute Petrosian-$r$ radius 
(in kpc) versus the redshift $z$ is displayed for HSB galaxies. 
The average of the distribution within redshift bins
of 0.01 is shown as red dots. The typical value for the statistical error
of these averages is 0.03\,kpc. Therewith it is smaller than the size of the 
points.
The lower right panel shows the same diagram as the lower left panel, but  for 
LSB galaxies. Error bars indicate the statistical error of the
individual average values. Mind, that with increasing $z$ the averaged absolute
Petrosian size of the galaxies also grows in both the HSB and LSB case.}
\label{petrosian_fig}
\end{center}
\end{figure*}

It is conspicuous that for the redshift interval of $0.01<z<0.055$ 
the diagram shows a LSB galaxy population which is dominated by dwarf-like
galaxies. The mean value of the absolute magnitude distribution in this
redshift interval is 
$\overline{M}_B=-16.58\pm0.08$\,mag. 
The peak of the distribution lies at a value of $\sim-18.0$\,mag.  
These value places the dominant LSB galaxy population of the redshift interval 
of $0.01<z<0.055$ into a region of dwarf-like, irregular galaxies and
small spirals in the galaxy luminosity distribution. 
The average value is similar to the luminosity 
of the Small Magellanic Cloud ($M_B=-16.5$\,mag,
\citealt{van_den_bergh00}), the peak value is similar to that of the 
Large Magellanic Cloud ($M_B=-18.0$\,mag, \citealt{van_den_bergh00}).

For the higher redshift interval with 
$0.055<z<0.1$ the situation changes. The peak of the distribution migrates
towards the brighter region of the absolute magnitude diagram in comparison
to the other redshift interval. This means that there the LSB population is
dominated by larger, more massive galaxies. 
This is confirmed by the mean values of the distributions. 
A mean value of $\overline{M}_B=-18.13\pm0.10$\,mag was calculated. 
This value is similar to that of the Large Magellanic Cloud ($M_B=-18.0$\,mag,
\citealt{van_den_bergh00}). The peak of the distribution is located 
at a value of $\sim$-19.3\,mag (which is similar to that of M\,33).
This also places these
galaxies in the total magnitude range of HSB spiral galaxies.

The lack of low-luminosity LSB galaxies at the higher redshift interval is not 
caused by dimming effects since the Petrosian magnitude is free from 
cosmological dimming effects (Section \ref{LSB_selection}). 
It is obvious that this lack is
caused by the apparent Petrosian $r$-magnitude limit of
$r<17.77$\,mag. It causes 
smaller galaxies with a low absolute magnitude to be excluded from
spectroscopic targeting at the higher redshift interval.
For example, a galaxy with a total absolute $r$-magnitude of $M_r=-17.50$\,mag
at a redshift of $z=0.055$ has an apparent magnitude of $m_r$=19.30\,mag
and therefore it would be excluded from spectroscopic targeting by the
magnitude limit, unless the field is sparse populated so that
there are still fibers left after the application of the strict 
criterias for spectroscopic targets.

More puzzling than the absence of dwarfish galaxies at the higher redshift
interval of $0.055<z<0.1$ is the apparent
lack of large LSBs at lower redshifts 
of $0.01<z<0.055$. Indeed, it is not the case that there are no large
galaxies in the lower redshift interval of $0.01<z<0.055$, but they are 
overwhelmed by small galaxies. This effect is not only observed for LSBs
but also for HSB galaxies of the SDSS sample.
This is seen in Figure \ref{petrosian_fig}.

The upper left panel of the figure shows the angular 
Petrosian-$r$ radius in arcsec versus the redshift of the galaxy  
for all HSBs.
At the upper right panel, the same is shown for LSBs.
Red dots are the mean values of the apparent Petrosian-$r$ radius 
distribution within the corresponding redshift interval, in the upper left 
panel for HSBs and in the upper right panel for LSB galaxies. 
The binning of the redshift intervals is $\Delta z=0.01$.

The distribution of the average angular Petrosian-$r$ radius of both galaxy 
types (in both upper panels) is quite flat. For HSBs (left upper panel)
the average apparent Petrosian-$r$ radius declines by $\sim60\%$ between 
the redshift of $z=0.01$ and $z=0.1$,
whereas the comoving distance (the distance obtained from redshift by
assuming Hubble flow) is increased tenfold. 

A similar situation is found in the LSB data. The averaged distribution 
of the angular Petrosian radius (red dots in the right upper panel)
shows the same declining trend as for HSBs between a redshift of 
$z=0.01$ and $z=0.1$ but the average apparent radius is only decreased by
$\sim40\%$ within that redshift range. In both diagrams there are three
horizontal line-like structures. These lines are artefacts due to
a bug of the SDSS pipeline, since it was checked by eye that all objects 
forming a line in that diagram do definitively not have the same angular size.
The problem is known by the SDSS pipeline programmers and it is
caused by the ``mismatches between the spectroscopic and imaging data''.
On the DR4 homepage it is stated that for various reasons, a small fraction of 
the spectroscopic objects do not have a counterpart in the best object 
catalogs. In addition, the DR4 does not contain photometric information for 
some of the special plates, and the retrieval of photometric data from the CAS 
database requires special care for objects from the special plates 
(see also: www.sdss.org/dr4/start/aboutdr4.html).

The lower panels show the Petrosian radii in kpc calculated from the
angular size and the comoving distance obtained from redshift (left panel
for HSBs, right panel for LSBs). Since the
data are not corrected for Virgocentric infall the comoving distances and 
therewith the calculated radii are accurate only for redshifts beyond
$z=0.01$. For HSBs the average apparent Petrosian radius decreases by
a factor of $\sim2.5$ between redshift $z=0.01$ and $z=0.1$, but the redshift
increases by a factor of ten, the average absolute Petrosian radius also
increases. It does not grow linearly  with a factor of four but with a factor
of 2.5, since the comoving
distance is not directly linear with $z$ due to relativistic corrections.  
The situation for LSBs (lower right panel) is similar, again. The 
averaged absolute Petrosian radius also rises between a redshift of
$z=0.01$ and $z=0.1$ with a slope of $\sim4$. This is caused by a
decreasing angular radius with a slope of about 5/3 within
a redshift range from $z=0.01$ to $z=0.1$ but a comoving distance 
increased by a factor of ten, minus relativistic corrections.   
Both the HSB and LSB lower diagram also show artefacts which are 
the same artefacts as in the upper panels but now producing diagonal lines
due to the calculation of the absolute Petrosian radii by calculating the 
comoving distances from the redshifts. 

This effect seen in the diagrams of Figure \ref{petrosian_fig}, that
with increasing distance on average larger galaxies are sampled,
is mainly caused by the apparent magnitude limit for 
spectroscopic follow up of Petrosian $m_r\leq17.77$\,mag. For
reaching this magnitude limit galaxies at higher 
distances must of cause be larger and therewith more luminous in 
total absolute magnitude.

The fact that the sample LSBs must be larger than their HSB 
equivalents to reach this limit due to their low stellar surface 
densities is obvious. From this it is clear that at lower redshifts,
both the LSB and the HSB sample are dominated by dwarfish galaxies, but 
at higher redshifts these galaxies do not get over the magnitude limit.
The   $m_r\leq17.77$\,mag limit is not
a sharp criterion, because in case that not all fibers are used for
bright galaxies free fibers are assigned to galaxies which are below
that limit. However, this case is rare.
Nevertheless, a sparse population of dwarfish galaxies is sampled at
higher redshifts as well. 

After understanding the SDSS selection function we were able to start 
examinating the environment of LSB galaxies.
Furthermore, the selection function gives the possibility 
to switch the probed LSB galaxy population between a sample consisting of 
large LSBs and another sample dominated by smaller LSBs just by changing the 
redshift interval. 
Moreover, the HSB comparison sample can also be changed between
a sample also containing smaller galaxies and a sample without 
small galaxies.

%
\section{Results: The Environment of LSBs} 
The data for the environment studies were obtained from the main galaxy
sample of the SDSS DR4.  
As described above, a redshift limit of $z\leq0.11$ was applied. Thereby, 
a sufficient accuracy for the redshift of 30\,km/s was achieved by
demanding a $z$-confidence of more than 90\%.   
Galaxies with a central, Tolman-dimming corrected surface
brightness of $\mu_B\geq22.5$\,mag/arcsec$^2$ were flagged as LSB galaxies,
otherwise they were flagged as HSBs.
Thereby the central $B$-surface brightness
was obtained by fitting exponential profiles (equation \ref{mue_profile}) to 
the azimuthally averaged 
surface brightness profiles measured by the SDSS pipelines and the
central surface brightnesses in the SDSS modified Gunn bands $g$ and $r$ 
which were converted into a $B$-surface brightness using the transformation 
equations of \cite{smith02}.

The uncertainty of the central surface brightness obtained from fitting surface
brightness profiles to the surface-brightness distribution of each galaxy 
produces false classified galaxies in both the HSB and LSB bin. To 
estimate the number of false classified HSBs that contaminate the LBS bin 
Monte-Carlo simulations on the surface-brightness distribution were performed. 
Therefore the obtained distribution in central surface brightness of all 
galaxies was convolved with an Gaussian distribution representing the 
uncertainty in the central surface brightness.   
With a typical value of about $0.1$~mag/arcsec$^2$ for this uncertainty 
(standard deviation) in the central surface brightness 
the amount of intrinsically HSBs pushed into the LSB bin spuriously and vice 
versa was determined. 
We found the number of HSBs spuriously classified as LSBs to be 13\% of the 
total amount of galaxies in the LSB bin. Therefore the LSB sample is 
contaminated with 13\% of falsly classified HSBs.
Coevally, the simulation shows that 5\% of true LSBs are shifted into the HSB 
bin due to this uncertainty in central surface brightness.  
Therefore we conclude the LSB sample to contain an adequate number of true LSBs
for performing environment studies on LSBs as described in the following.

In order to probe the environment of the LSB galaxies one has to count
neighbor galaxies using a certain search volume. It is clear, that as 
a neighbor both LSB and HSB galaxies count. Additionally, the galaxy density
in the vicinity of LSBs was compared to the galaxy density in the vicinity of
HSBs. For these two reasons, the LSB and HSB galaxies were stored into one
file, but with different flags indicating LSB or HSB property. Hence,
and with an appropriate neighborhood analysis code, it
was guaranteed that as neighbor of a scrutinized LSB galaxy (and of course a 
HSB) both LSB and HSB galaxies count for indicating the surrounding galaxy 
density.
\subsection{The Pie Slice}\label{section_pie}

\begin{figure*}[t!]
\begin{center}
\includegraphics[width=8.7cm]{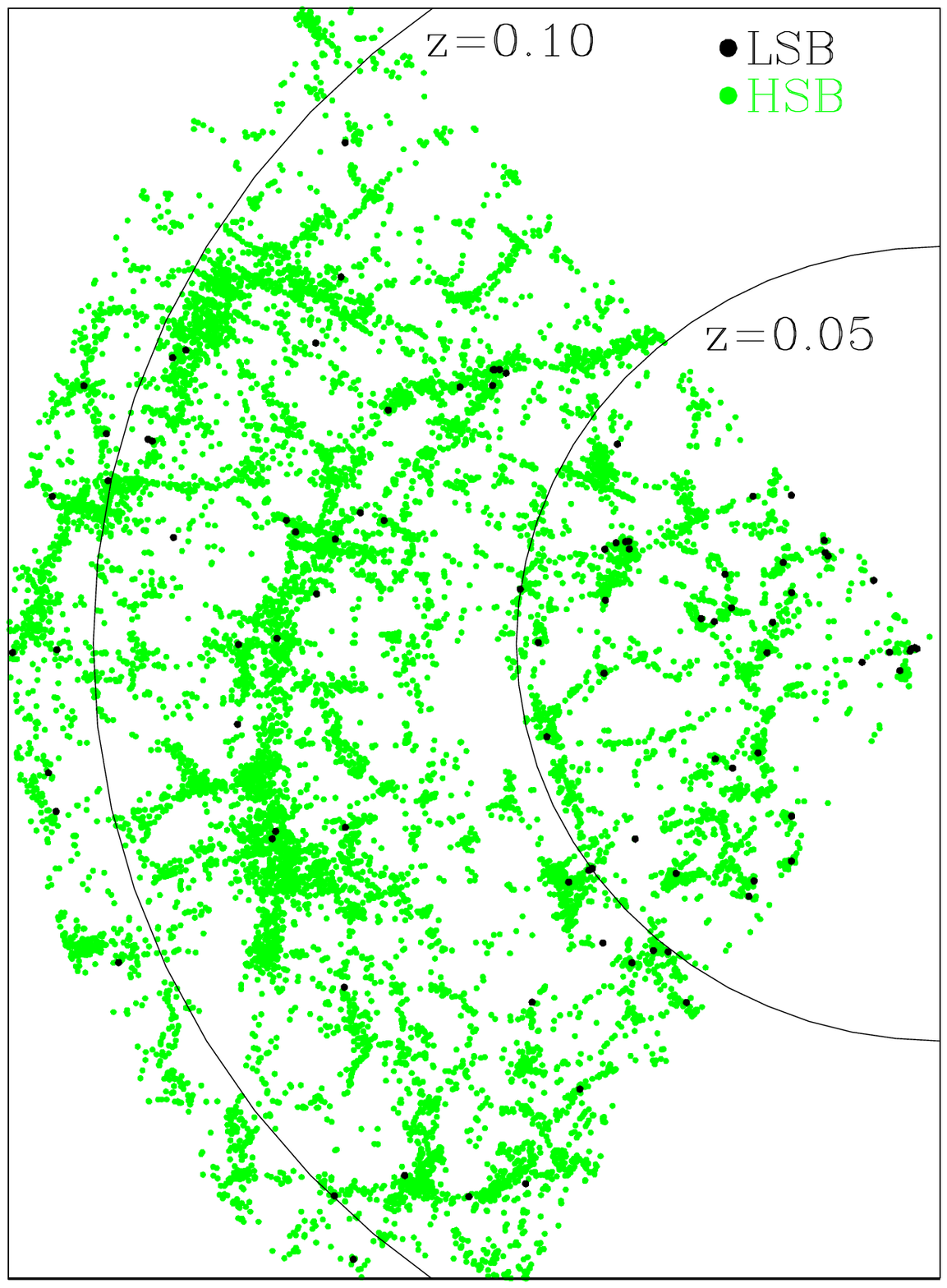}
\includegraphics[width=8.65cm]{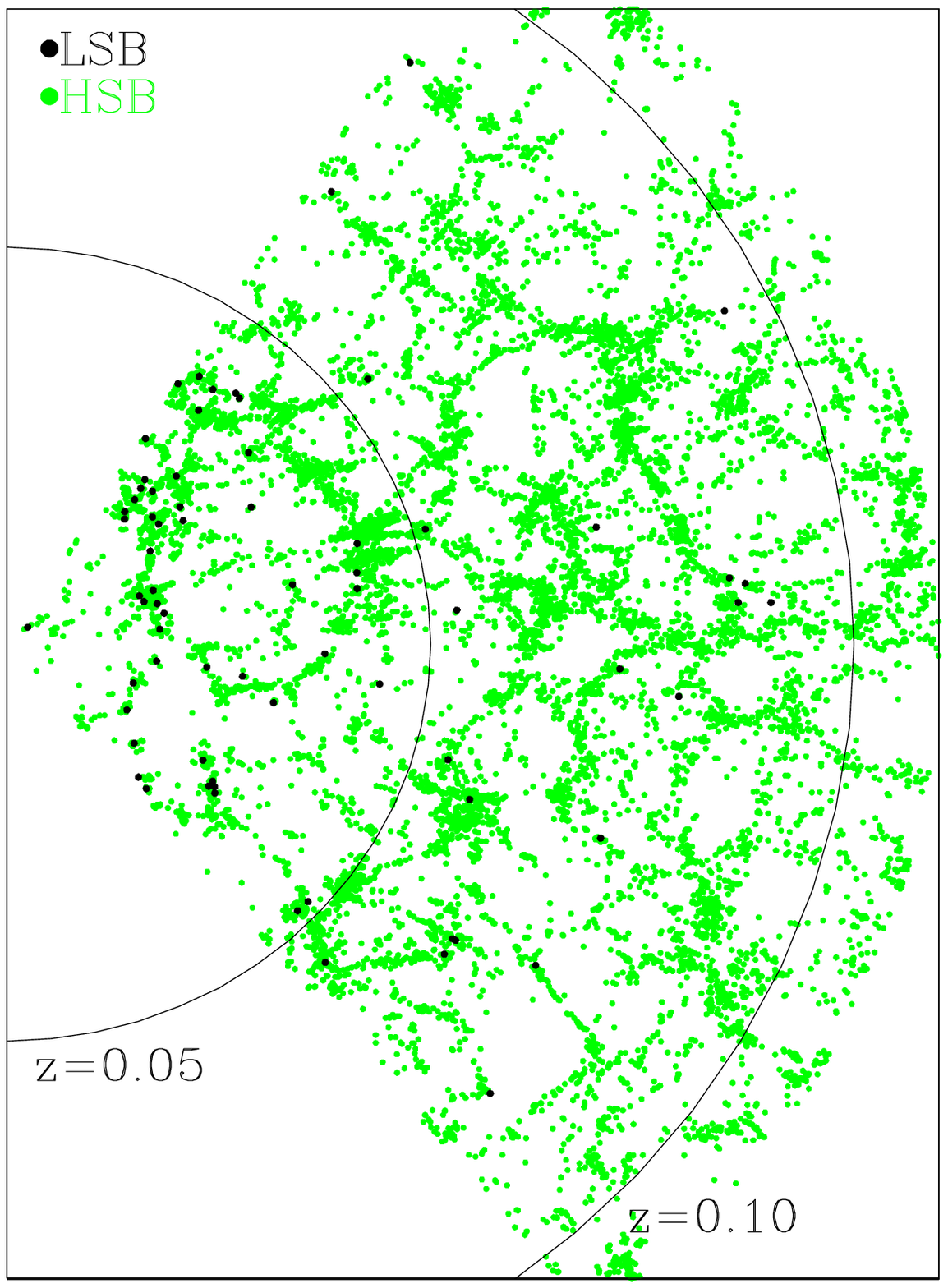}
\caption[Pie Slice Diagrams]{Two analyzed pie slices produced from
SDSS DR4. Black dots are LSB galaxies and green dots represent HSBs.
Left panel shows the distribution of LSBs and HSBs within a right ascension 
range of 120\degr$\leq\alpha_{2000}\leq 240$\degr~and a redshift of 
$z\leq 0.11$ in a polar plot. The declination range of 
-1.25\degr$\leq\delta_{2000}\leq 1.25$\degr~is projected onto the plane.
Right panel shows the same, but for a right ascension range 
of 310\degr$\leq \alpha_{2000}\leq 360$\degr~ and 
0\degr$<\alpha_{2000}\leq 60$\degr. The declination range is again 
-1.25\degr$\leq\delta_{2000}\leq 1.25$\degr. Mind, that within a redshift interval 
of $0.05\leq z\leq0.11$ the LSBs are mainly located at the outer parts of the 
filaments and walls of the LSS defined by HSBs. However, for 
a redshift of $z<0.05$ the situation is not so clear. There one can find
LSB galaxies at the outer parts of the filaments as well as in the middle
of walls and clusters.
Thereby, one has to take into account that at the lower redshift interval 
the LSB population is dominated by dwarfish galaxies and at higher redshifts 
the LSB sample mainly consists of larger galaxies.}
\label{pie_slice}
\end{center}
\end{figure*}
For a first glance at the distribution of the LSB galaxies within the 
large-scale structure (LSS), so called pie slices were plotted from the
database obtained as described in the section before. Figure \ref{pie_slice}
shows such pie slice diagrams, where the distribution of right ascension and 
the redshift of LSB (black dots) and HSB galaxies (green dots) are 
displayed in polar plots. The left panel contains the right ascension range 
which was taken from an equatorial scan region of the DR4 with 
120\degr$\leq\alpha_{2000}\leq 240$\degr~ 
 and the declination range of -1.25\degr$\leq\delta\leq 1.25$\degr~ is 
projected onto the plane, whereas the redshift is limited to a value of 
$z\leq0.11$. The right panel shows a pie slice of the same declination range,
but with a right ascension of 310\degr$\leq \alpha_{2000}\leq 360$\degr~ and 
0\degr$<\alpha_{2000}\leq 60$\degr.
The left panel contains 94 LSBs and 12768 HSBs, the numbers for the right panel
are 72 LSBs and 11379 HSBs.


\subsection{Neighbor Counting within Spheres}
In order to quantify the differences in the surrounding galaxy densities of 
LSBs with respect to that of HSBs statistical neighbor counting was performed.

\subsubsection{The Neighbor Counting Algorithm}
For this analysis, an algorithm for counting neighbors was developed.
This algorithm works as follows. For each galaxy,
a sphere with a certain radius is defined with the probed galaxy 
in the center. Then the number of neighbor galaxies within this sphere is 
counted. This step is performed for all galaxies found in the 
program input file.
Since the data of LSB and HSB galaxies are stored in one file which 
is used as the data input for the program, both galaxy types count as 
neighbors independent from if the scrutinized galaxy is a LSB or HSB.
The radius of the sphere is an input parameter which the user is asked to
define at the program start, as well as the names of the input and output
files. After the interrogation of the input parameter and files,
the program calculates for each galaxy of the input file the comoving 
distances by applying a Hubble constant of 71\,kms$^{-1}$Mpc$^{-1}$ 
\citep{spergel03} and the speed of light. 
Thereby,  
relativistic corrections for the redshifts were used. Since the
environment studies were limited to a reshift range of $0.01\leq z\leq0.1$, 
neither Virgocentric infall was corrected nor more complicated
streaming motions than pure Hubble flow 
was taken into account.  
After that the right ascension,
declination, and comoving distances are converted into Carthesian coordinates.
Then the code starts neighbor counting within the 3-dimensional distribution 
of LSB and HSB galaxies by centering a sphere with a radius which was 
specified at program startup  on each galaxy of 
the input file and then counting the neighboring galaxies within this  
sphere. 

In order not to distort the statistical results at the borders
of the catalogue volume, an edge correction was applied.
The borders of the sample
were avoided so that galaxies whose spheres were cutting the edges of
the sample volume were rejected and not stored in the output file. 
Since all galaxies in the input file are HSB or LSB type flagged, 
one has the possibility to divide the result into statistics for
the environment of LSBs and HSBs separately. For edge correction the covered
volume of the input catalog was sampled with cubes of different sizes which 
did not cut the borders of the catalog. 

Due to the fact that at lower redshifts ($0.01\leq z\leq0.055$) the
sample LSBs are dominated by dwarfish galaxies and at higher redshifts
($0.055\leq z\leq0.1$) mostly large galaxies are contained in the LSB sample, 
the environment study had to be
separated into these two redshift intervals. 
For each redshift interval, the environment studies were performed in 
different runs with several values for the sphere radius. 
It was varied from 0.8\,Mpc to 8\,Mpc in steps of
0.6\,Mpc. The lower border of the scale range (0.8\,Mpc) was chosen 
to avoid bias effects 
to the statistics caused by fiber placement constraints, since
the minimum possible distance between two adjacent fibers is 55'' in
angular distance. This value
corresponds to a minimum distance between two adjacent galaxies of 0.112\,Mpc 
at a redshift of $z=0.1$ for getting spectra of both galaxies.
Hence, with a sphere radius of  $r=0.8$\,Mpc, effects caused by
fiber placement constraints are well sampled.

 The upper border of the scale range (8\,Mpc) of the 
neighbor counting was chosen, because with higher values the probed
volume decreases due to boundary corrections and the statistics drops
in significance. However, the chosen scale range is sufficient to probe
the spatial distribution of LSBs on group radius scales (1-3\,Mpc, 
\citealt{cox00, krusch03})
and filament sizes ($\sim 5$\,Mpc, e.g. \citealt{white87, doroshkevich97}).

\subsubsection{The Galaxy Cluster Finding Algorithm}
\label{cluster_finding}
\begin{figure*}[t!]
\begin{center}
\includegraphics[width=17.0cm]{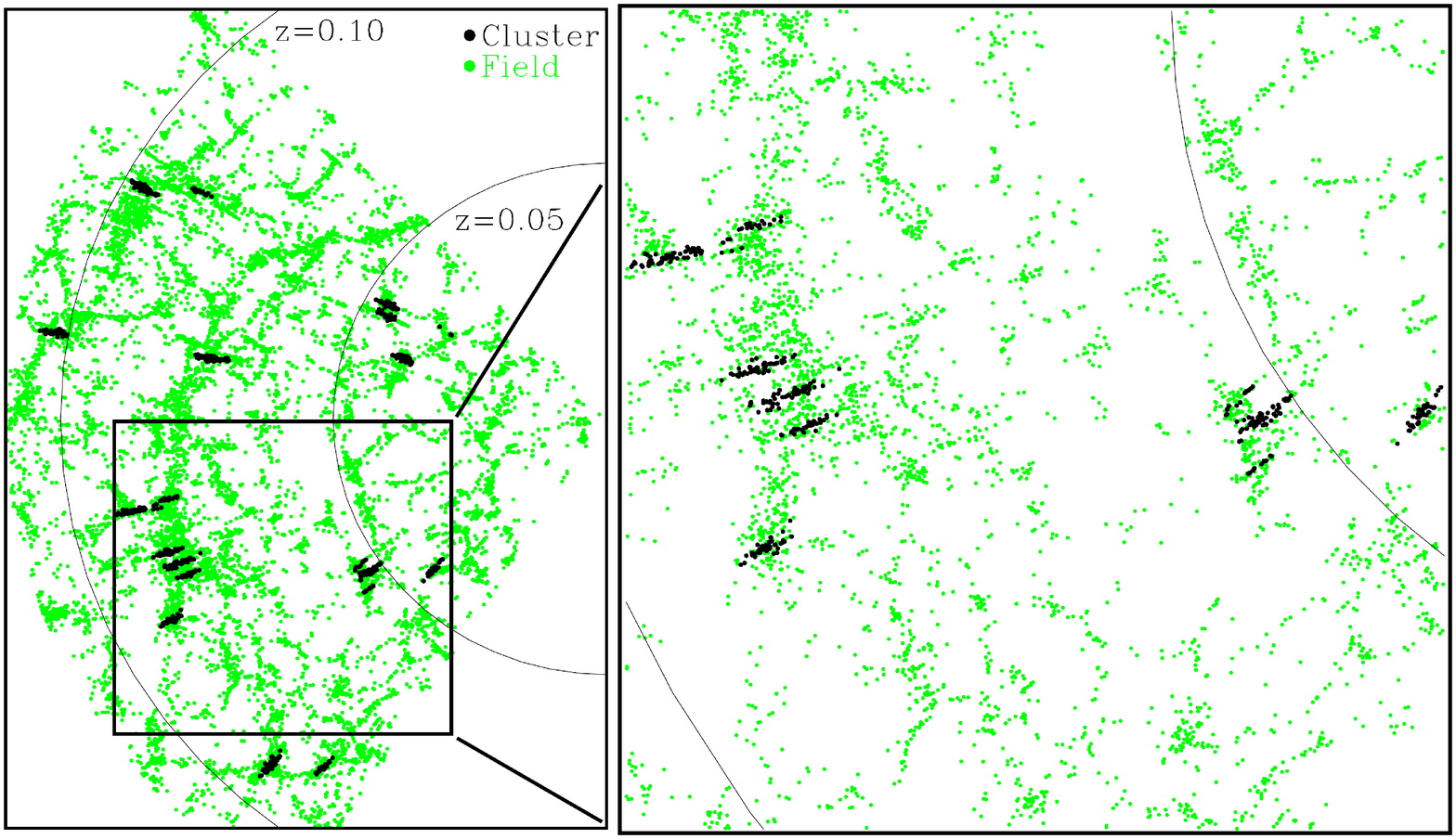}
\caption[Cluster Finding Algorithm]{Pie slice to demonstrate the
cluster finding algorithm. The left diagram shows the distribution 
of all HSB galaxies within a right ascension range of 
120\degr$\leq\alpha_{2000}\leq 240$\degr~and a redshift of $z\leq 0.11$ in a 
polar plot. The declination range of 
-1.25\degr$\leq\delta_{2000}\leq 1.25$\degr~is projected onto the plane.
The galaxies found to be arranged in clusters detected by the 
cluster finding algorithm are marked with black dots,
whereas the field galaxies are displayed as green dots. It can be seen, that
the cluster searching code finds the ``finger of god''-like structures in the
LSS. Additionally, these plots show that all clusters are embedded in
wall-like structures of the LSS which indicates that the structure formation
has not been completed, yet.}
\label{cluster_pie}
\end{center}
\end{figure*}

After the analysis using the code for neighbor counting,
cluster galaxies had to be excluded from the sample in a second statistical
analysis for a comparison
of the clustering properties of field LSBs and HSBs. 
This was done due
to the fact that, when calculating comoving distances from redshifts,
the velocity dispersion (which can exceed values of $\sigma\geq 1000$\,km/s)
of galaxy clusters mocks an extension of the cluster 
in the line of sight direction. These structures bias the results in
the LSS and had to be removed from the data. Therefore, an algorithm
for finding clusters was developed.

This algorithm searches for clusters
in the LSS by counting galaxies within a cylinder aligned radially
in the line of sight direction
with a configurable radius and height. The radius $r$ of the cylinder
thereby complies with the radius of the cluster and the height of the
cylinder corresponds to its velocity dispersion $\sigma$. These two values are
parameters which can be chosen by the user of the program.
The minimum number of cluster members ($N_{members}$) can also be set at 
program startup.

In a parameter study, the best
parameters concerning $r$, $\sigma$, and $N_{members}$ for finding galaxy 
clusters were figured out.  It turned out that,
the best results of the hunt for galaxy clusters producing ``fingers of
god'' were achieved with the parameters cylinder radius $r=2.5$\,Mpc, 
velocity dispersion $\sigma=1000$\,km/s and the minimum number of
cluster galaxies $N_{members}=50$. 
Our cluster search code is based on a definition similar to
the definition used by \cite{abell58} and \cite{abell89}, but it works in
three dimensions. 
They
defined the galaxy clusters mainly by a richness and a compactness criterion. 
The richness criterion requires that, a cluster must contain at least 
50 members that are not more than 2\,mag fainter than the third brightest
member. In the compactness criterion it is demanded that a cluster must be
sufficiently compact that its 50 or more members are within a given
radial distance $r$ of its center.

The cluster finding algorithm was applied to the catalog containing the DR4 
LSB and HSB sample after the neighbor counting within spheres. 
The program produced a file containing all (LSB and HSB) galaxies of
the input files except the cluster galaxies of the clusters found
by the program. This means that the cluster galaxies were removed from
the galaxy sample file (containing neighborhood informations) retroactively, 
which means that only the distribution of  field galaxies flows into the 
statistic.
Otherwise, if one performs the cluster removal before environment analysis,
one would produce holes into the large-scale structure which had to
be masked out before the environment analysis. 
If not masked out, they would distort
the statistics like boundary effects (which were also excluded).

\subsubsection{The resulting LSB Environment}\label{environment}
 
\begin{figure*}[t!]
\begin{center}
\includegraphics[width=8.5cm]{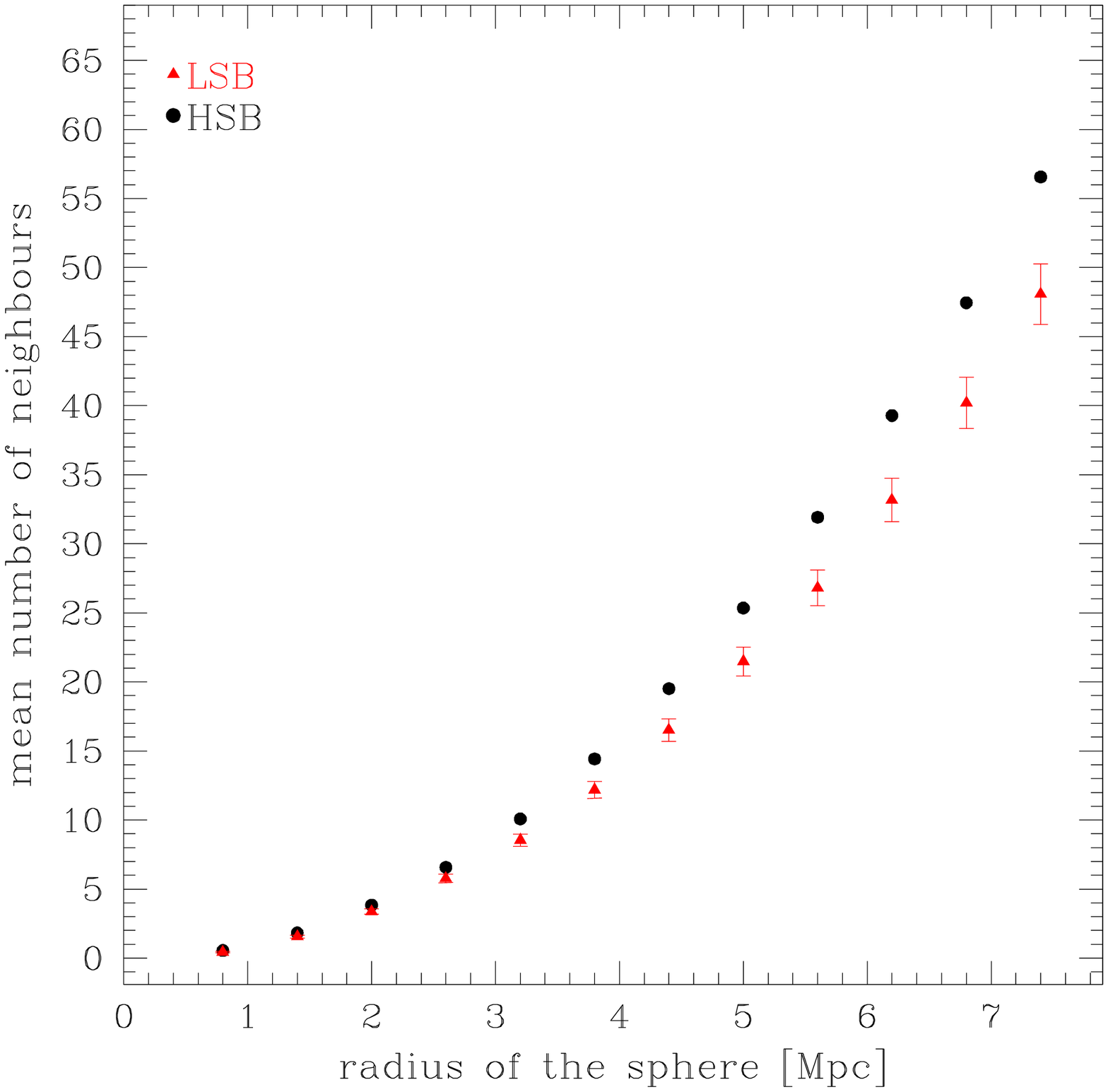}
\includegraphics[width=8.5cm]{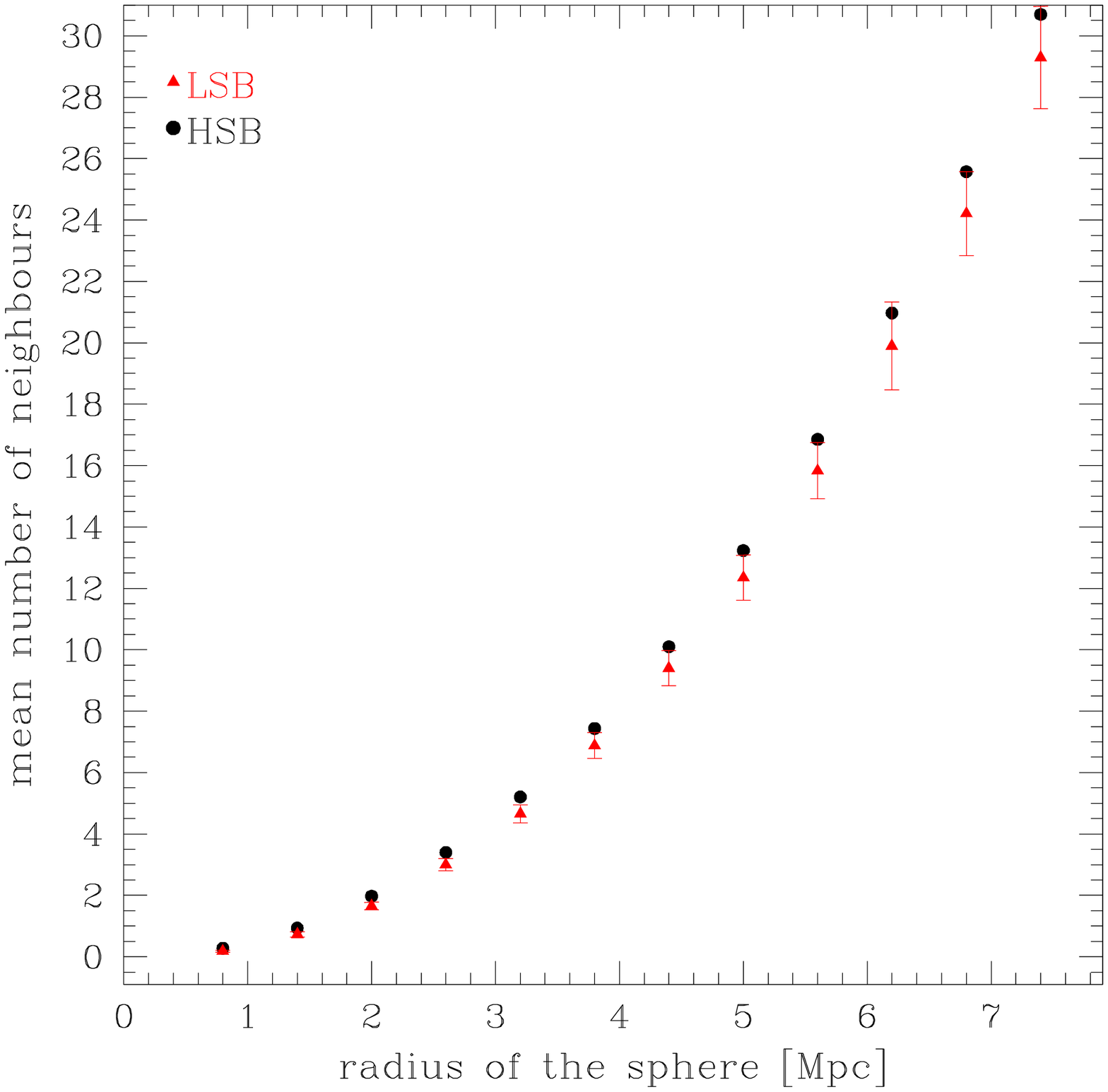}
\caption[Mean Neighbors with Clusters]{The diagrams show the average number
of neighbors for LSBs (red triangles) and HSBs (black dots) versus the
radius of the sphere within which the neighbors were counted. 
Both plots are not corrected for cluster galaxies.
Left diagram
displays the results of the 
redshift interval of $0.01\leq z \leq 0.055$, right panel shows
the same but for the redshift range of $0.055\leq z \leq 0.1$. 
The average number of neighbors stays for LSBs systematically below
the values of the HSB statistics both for the lower redshift interval (left)
and the higher redshift interval (right).}\label{mean_neighbors_uncleaned}
\end{center}
\end{figure*}
During several runs of the neighbor counting algorithms as described 
above, the radius of the sphere was varied between 0.8\,Mpc and 8.0\,Mpc
in steps of 0.6\,Mpc. 
The number of neighbors of each individual run of the program (with a 
fixed sphere radius) were averaged for LSB
and HSB galaxies. Because of the SDSS selection function the sample 
contains LSB galaxies of different size and total luminosity at different 
redshifts, namely the low redshift interval of $0.01\leq z \leq 0.055$ 
is dominated by dwarfish LSBs. The higher  redshift range of 
$0.055\leq z \leq 0.1$ contains mainly large LSBs.
Therefore the examinations on the environment were divided into two
symmetric redshift bins corresponding to these redshift intervals. 

\begin{figure*}[t!]
\begin{center}
\includegraphics[width=8.5cm]{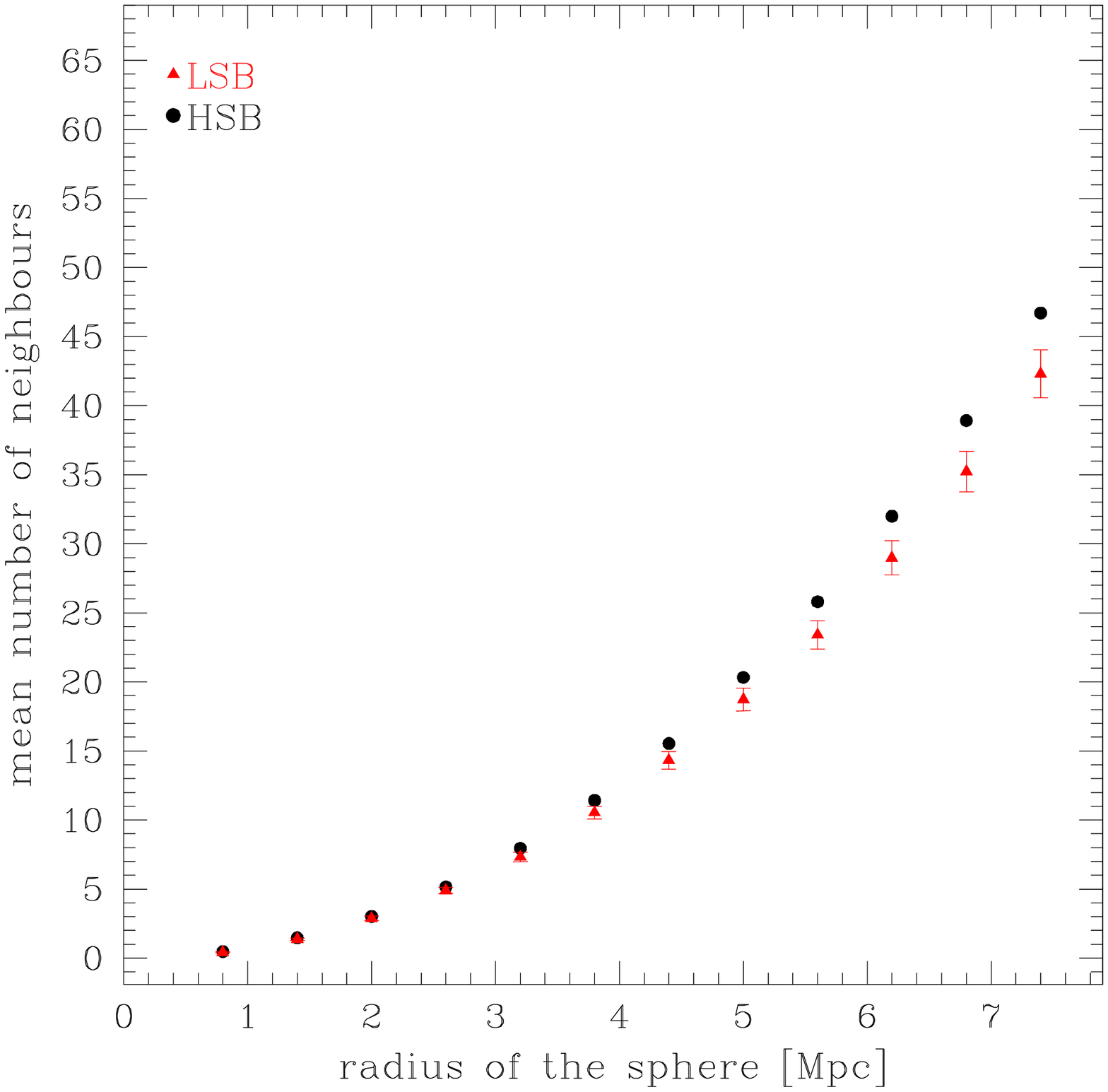}
\includegraphics[width=8.5cm]{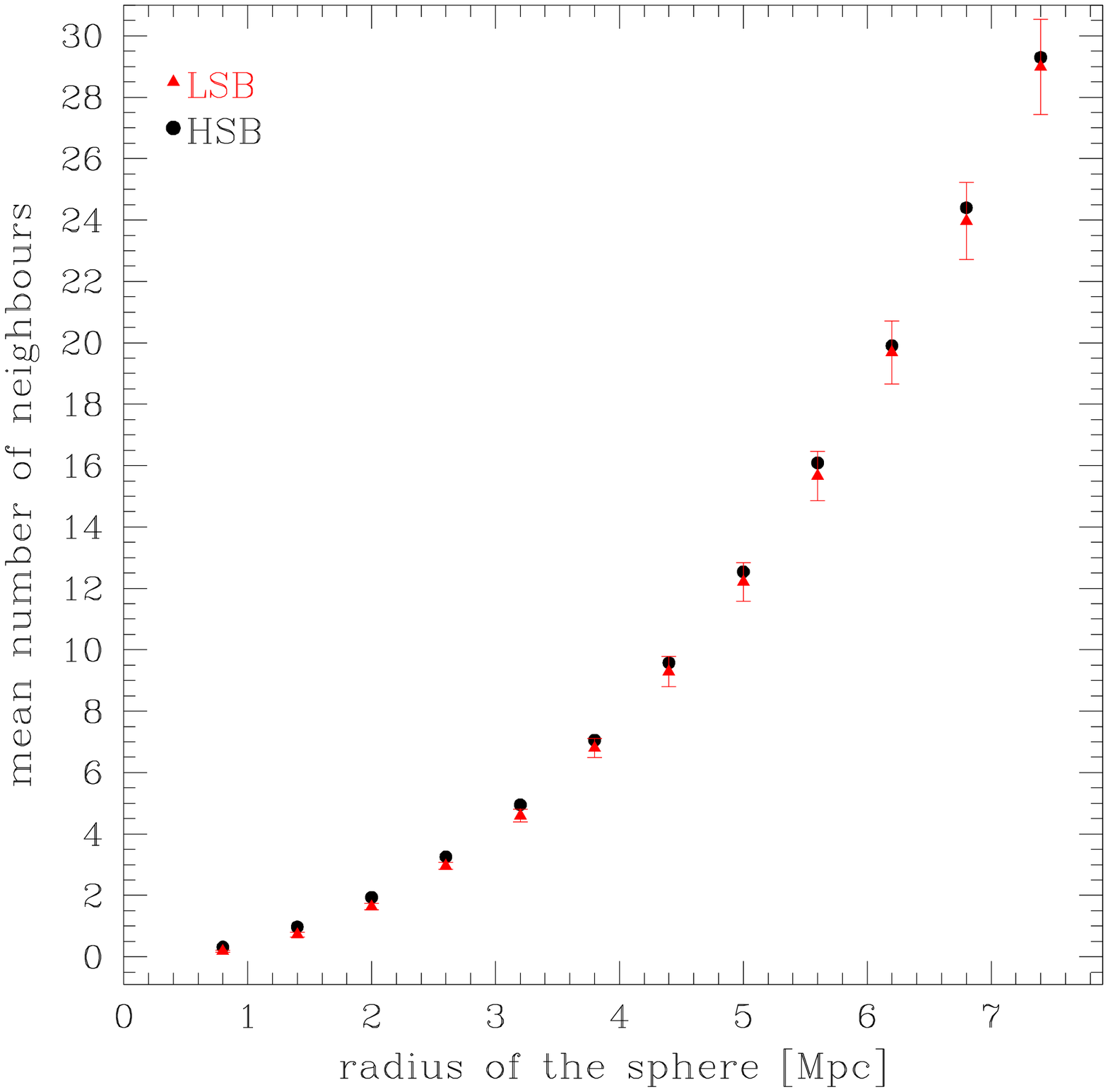}
\caption[Mean Neighbors without Clusters]{The panels show again the average 
number of neighbors for LSBs (red triangles) and HSBs (black dots) versus the
sphere radius within the two redshift intervals (left: $0.01\leq z \leq 0.055$,
 right: $0.055\leq z \leq 0.1$), but at this time the cluster galaxies were
removed from the statistics (as described in section \ref{cluster_finding}). 
This means that the distribution of LSBs in comparison to HSBs was probed in 
the field. On average the LSB galaxies still have less neighbors than HSBs
but the signal is not so strong (but still significant)
as if we average over cluster and field galaxies 
(Figure \ref{mean_neighbors_uncleaned}).}
\label{mean_neighbors_cleaned}
\end{center}
\end{figure*}

Figure \ref{mean_neighbors_uncleaned} shows the average number of
neighbors for LSBs (red) and HSBs (black) versus the sphere radii
within the redshift intervals $0.01\leq z \leq 0.055$ (left panel) and
$0.055\leq z \leq 0.1$ (right panel). To produce this diagram
neighbor counting was performed in several runs within spheres with
radii between 0.8\,Mpc and 8\,Mpc in steps of 0.6\,Mpc. For each sphere
radius the number of neighbors was averaged for the LSB galaxies
and for the HSB galaxies (as neighbor both galaxy types counted).
In the lower redshift interval
$\sim400$ LSBs were probed in comparison to $\sim31000$ HSBs.
For the higher redshift interval the sample contains $\sim200$ LSB and 
$\sim69000$ HSB galaxies. In this case, cluster correction was not applied to 
the data. The data show that LSB galaxies have on average less
neighbors than HSB galaxies on scales between 0.8\,Mpc and 8.0\,Mpc. 
This is the case for both redshift intervals (although the trend is less 
significant in the higher redshift interval).
This means that dwarfish LSBs as well as large LSBs are preferably found 
in regions with lower galaxy density than in the vicinity of HSBs.
The next step was to probe the location of LSBs in the LSS without clusters.
For that, the cluster galaxies
were removed from the statistics using the cluster finding algorithm 
(section \ref{cluster_finding}). This means, that all galaxies, which
are located in a LSS volume occupied by a cluster, were removed from the
statistics that it was averaged over pure field galaxies.
Figure \ref{mean_neighbors_cleaned} shows the results of that study.
Again, two redshift bins (left panel: $0.01\leq z \leq 0.055$, right: 
$0.055\leq z \leq 0.1$) were examined. 
Galaxies which have more than
50 neighbors within a cylinder with a radius of 3\,Mpc and a height of
1000\,km/s aligned with its axis towards the line of sight were rejected.  
Again, the number of neighbors were averaged for LSBs and HSBs at different
sphere radii. The diagrams show that on average LSB galaxies have less
neighbors than HSB galaxies on all probed scales for both dwarfish
and large LSB galaxies, since all triangles representing the 
average LSB number of neighbors are located systematically below 
the corresponding averaged values for HSBs. For the redshift bin with
$0.01\leq z \leq 0.055$ (left panel) it becomes statistically significant at a
sphere radius of 3.2\,Mpc. For the higher redshift bin (with 
$0.055\leq z \leq 0.1$, right panel) it is statistically significant between
2\,Mpc and 3.8\,Mpc. As all LSB values are located below the HSB values, one
can argue that the effect is also seen on scales with lower statistical
significance,
since the probability is small that all LSB points are located below the HSB values due to statistical noise.  
                 
The statistical significance of the statement that LSB
galaxies have on average less neighbors than HSBs had to be probed.
For that, 
Kolmogorov-Smirnov  (KS, \citealt{chakravarti67}) two sample tests were
performed on the LSB and HSB distribution of
neighbors from which the average values were calculated.  
The KS test is intended to probe the null hypothesis that two data samples 
come from the same distribution. 
Since the 
KS statistic is used for unbinned data sets, it was applied to the unbinned
number of neighbor distribution of LSBs forming the first data sample
and the same distribution for HSBs producing the second test data sample.

The results are presented in Table   
 \ref{KS1} and \ref{KS2}. Thereby, Table \ref{KS1} refers to the significance
of the null hypothesis in Figure \ref{mean_neighbors_uncleaned}
with respect to the corresponding scale radii.
This Figure contains statistics including both cluster and field galaxies. 
For the lower redshift range with $0.01\leq z< 0.055$ 
(Figure \ref{mean_neighbors_uncleaned}, left diagram), our hypothesis
that LSB galaxies have on average less neighbors than HSBs holds with
more than $1\,\sigma$ probability for the scale range of 0.1\,Mpc to 2.6\,Mpc
and with around $2\,\sigma$ probability for scales between 3.2\,Mpc and 
5.0\,Mpc. For the scale interval of 5.6\,Mpc to 8.0\,Mpc, a significance
of around 3\,$\sigma$ is reached for this hypothesis. 

In this context, one should take into account the selection 
function for LSBs (Figure \ref{abs_mag_distrib}), which shows that this
scrutinized redshift range is dominated by dwarfish LSB galaxies. Then, this
result shows that there exists a density contrast for small LSBs, which are on 
average located in a less dense environment than 
field and cluster HSBs.
In diagram \ref{mean_neighbors_uncleaned} (right panel) this 
hypothesis holds with explicitly more than $1\,\sigma$ 
significance for values of the sphere radii between 0.8\,Mpc and 2.0\,Mpc.
For the scale values of 2.6\,Mpc and 3.2\,Mpc the probability of the
hypothesis is still around $1\,\sigma$, but it drops below the value
for higher scale radii. However, all LSB neighboring values in Figure 
\ref{mean_neighbors_uncleaned} (right) at that scales are located 
systematically below the average number of neighbors for HSBs. This
indicates that this effect is real with a higher probability and over
a larger range of radii than the KS test indicates.

\begin{table}[t!]
\begin{center}
\caption[Significance of Neighbor Counting I]
        {Significance of neighbor counting statistics 
} 
\label{KS1}
\begin{tabular}{|l|l|l|}\hline
\multicolumn{3}{|c|}{Significance of Statistics}\\\hline\hline
z-range / refers to Figure & radius [Mpc] & significance [\%]\\\hline\hline
$0.01\leq z <0.055$      & 0.8 & 91.38\\ 
Figure \ref{mean_neighbors_uncleaned} (left) & 1.4 & 90.71\\  
not cluster corrected & 2.0 & 74.23\\ 
                      & 2.6 & 69.11\\  
                      & 3.2 & 94.25\\  
                      & 3.8 & 91.16\\  
                      & 4.4 & 94.37\\  
                      & 5.0 & 97.88\\  
                      & 5.6 & 99.26\\  
                      & 6.2 & 99.01\\  
                      & 6.8 & 98.55\\  
                      & 7.4 & 99.93\\  
                      & 8.0 & 99.48\\\hline 
$0.055\leq z <0.1$ & 0.8 & 79.26\\      
Figure \ref{mean_neighbors_uncleaned} (right) & 1.4 & 91.51\\     
not cluster corrected & 2.0 & 79.77\\      
                     & 2.6 & 64.48\\      
                     & 3.2 & 64.66\\      
                     & 3.8 & 52.17\\      
                     & 4.4 & 47.61\\      
                     & 5.0 & 55.95\\      
                     & 5.6 & 57.57\\      
                     & 6.2 & 34.50\\      
                     & 6.8 & 40.86\\      
                     & 7.4 & 44.38\\      
                     & 8.0 & 33.27\\\hline
\end{tabular}
\end{center}  
\end{table}

\begin{table}[t!]
\begin{center}
\caption[Significance of Neighbor Counting II]
        {Significance of neighbor counting statistics continued 
} 
\label{KS2}
\begin{tabular}{|l|l|l|}\hline
\multicolumn{3}{|c|}{Significance of Statistics}\\\hline\hline
z-range / refers to Figure & radius [Mpc] & significance [\%]\\\hline\hline
$0.01\leq z <0.055$   & 0.8 & 70.22\\ 
Figure \ref{mean_neighbors_cleaned} (left)    & 1.4 & 69.24\\  
cluster corrected     & 2.0 & 49.12\\  
                      & 2.6 & 56.81\\  
                      & 3.2 & 84.08\\  
                      & 3.8 & 77.84\\  
                      & 4.4 & 87.88\\  
                      & 5.0 & 95.06\\  
                      & 5.6 & 98.18\\  
                      & 6.2 & 97.63\\  
                      & 6.8 & 96.73\\  
                      & 7.4 & 97.39\\  
                      & 8.0 & 98.67\\\hline  

$0.055\leq z <0.1$   & 0.8 & 68.80\\    
Figure \ref{mean_neighbors_cleaned} (right)  & 1.4 & 86.59\\              
cluster corrected    & 2.0 & 64.36\\           
                     & 2.6 & 41.88\\           
                     & 3.2 & 41.57\\           
                     & 3.8 & 27.71\\           
                     & 4.4 & 37.05\\           
                     & 5.0 & 35.56\\           
                     & 5.6 & 38.55\\               
                     & 6.2 & 16.07\\               
                     & 6.8 & 21.55\\               
                     & 7.4 & 23.97\\               
                     & 8.0 & 16.41\\\hline         

\end{tabular}
\end{center}  
\end{table}

Table \ref{KS2} refers to the Figure \ref{mean_neighbors_cleaned} and
gives probabilities that the (inverted) null hypothesis that LSB galaxies have 
less neighbors than field HSBs is true on different scales for two redshift
intervals. Figure \ref{mean_neighbors_cleaned} contains the comparison
between field LSB galaxies and field HSBs, since all cluster galaxies were
removed from statistics. 

For the redshift range of $0.01\leq z <0.055$ the following
situation turns out to be as follows. For the scale values of 0.8\,Mpc and 
1.4\,Mpc this hypothesis holds with more than $1\,\sigma$ significance.
The sphere radii of 2.0\,Mpc and 2.6\,Mpc have a probability of around 50\%
for the trueness of the hypothesis. On scales between 3.2\,Mpc and
 4.4\,Mpc the statement possess clearly more than  $1\,\sigma$ probability.
And for the sphere range of 5.0\,Mpc to 8.0\,Mpc the significance 
is with values of $2\,\sigma$ up to $3\,\sigma$ quite high. 

For the higher redshift interval with $0.055\leq z <0.1$ the situation
is not clear. The first three scale values with 0.8\,Mpc, 1.4\,Mpc and
2.0\,Mpc have a significance for the hypothesis of around and about 
$1\,\sigma$. However, the probability that the hypothesis holds at higher 
sphere radii is low, below 50\%. 
Since the number of neighbors is on average still lower
for LSBs than for HSBs, this might be a hint that a difference in the LSB and HSB environment might exist in this redshift bin as well. 
However, when considering each value in diagram  
\ref{mean_neighbors_cleaned} (right) individually, the density 
contrast is not significant on scales of 2.6\,Mpc and above. 
This is an effect
of small numbers, which will be improved with further data sets 
covering larger sky areas.

\subsection{LSB-HSB Antibias}\label{antibias}
In order to find another independent 
method to quantify the clustering properties
of LSBs in comparison to HSBs, the LSB-HSB biasing parameter
was derived from the SDSS dataset. 
The galaxy bias is a term from cosmology and is normally used to describe
the difference in the clustering properties between galaxies and Dark
Matter on large scales. We use this stochastic 
bias measure for the first time to display differences in the 
clustering properties of LSBs in comparison to HSBs. 
The LSB-HSB Antibias parameter was derived from the Neighbor Counting
Analysis as described before by using the galaxy number density 
$N$ at the location of each galaxy within spheres of the radius $R$.
The second moment of the density contrast was then
calculated for the density field at the location of all sample LSBs and HSBs 
using the equations
\begin{eqnarray}
\label{eq_density_LSBs}
 <\delta_{R,\rm LSB}^2>\,=
\frac{1}{n}\sum_{i=1}^n\left(\frac{N_{i,R,\rm LSB}}{
\overline{N}_{R,\rm LSB}} -1\right)^2,
\end{eqnarray}
and  
\begin{eqnarray}
\label{eq_density_HSBs}
 <\delta_{R,\rm HSB}^2>\,=\frac{1}{n}\sum_{i=1}^n\left(\frac{N_{i,R,\rm HSB}}{
\overline{N}_{i,R,\rm HSB}} -1\right)^2.
\end{eqnarray}
 
Using these equations (\ref{eq_density_LSBs}, \ref{eq_density_HSBs}) in 
combination 
with the results of the environment studies within spheres of the radius $R$,
the average density contrast at the locations of LSBs and the
same quantity at the locations of HSBs were calculated.
Thereby, $N_{i,R,\rm LSB}$ is the number of (LSB and HSB) galaxies within a 
sphere of the radius $R$ centered on the LSB galaxy $i$ or expressed in a
different way the galaxy number density within a radius $R$ at the location of 
the $i^{\rm th}$ LSB galaxy. The quantity $n$ is the total number of 
LSBs in eq. \ref{eq_density_LSBs} (or HSBs in eq. \ref{eq_density_HSBs}) 
and $\overline{N}$ is the number of (LSB and HSB) galaxies within that 
radius $R$ averaged over all LSB galaxies.

From these results, the stochastic bias parameter $b(R)$ was obtained using 
the following equation 
\begin{eqnarray}
\label{eq_bias_LSB_HSB}
b(R)=\sqrt{\frac{<\delta_{R,\rm LSB}^2>}{<\delta_{R,\rm HSB}^2>}},
\end{eqnarray}
with $R$ the sphere radius, $<\delta_{R,\rm LSB}^2>$ the second moment
of the LSB density contrast in dependence of the radius of the probing
sphere and $<\delta_{R,\rm HSB}^2>$ the same but for the sample HSB galaxies.  

With the definition of this parameter it is now possible to express
the differences in the LSB and HSB environment in terms of the 
density contrast and galaxy bias.

\subsubsection{Results on the LSB-HSB Antibias}
The results obtained from probing the LSB-HSB galaxy bias 
are presented in 
Table \ref{bias}. In that Table, the values for the second moment of the 
density contrast for LSBs calculated using \ref{eq_density_LSBs}, for HSBs
(equation \ref{eq_density_HSBs}), and the resulting values for the LSB-HSB
bias parameter calculated from equation \ref{eq_bias_LSB_HSB} are given.
These values are displayed in dependence on the sphere radius. 
The results are again divided into the two redshift bins as used several times 
before. Furthermore, a division of the results into the cases 
``not cluster corrected'' and ``cluster corrected'' is done. The first case 
probes
the LSB galaxy bias with respect to all (field and cluster) HSBs. The second 
case tests if the distribution of LSB galaxies is biased against field LSBs.   

\begin{table}[t!]
\begin{center}
\caption[LSB-HSB Galaxy Bias Parameter]{LSB-HSB galaxy bias parameter 
}
\label{bias}
\begin{tabular}{|l|l|l|l|}\hline\hline
\multicolumn{4}{|c|}{LSB-HSB Galaxy Bias}\\\hline\hline
\multicolumn{4}{|c|}{redshift: $0.01\leq z<0.055$ / not cluster corrected}\\\hline
 $R$\,[Mpc] & $<\delta_{R,\rm LSB}^2>$ & $<\delta_{R,\rm HSB}^2>$ & $b(R)$\\\hline
  8.0 & 0.798 & 1.039 & 0.876\\ 
  5.6 & 0.912 & 1.256 & 0.852\\
  3.2 & 1.081 & 1.460 & 0.861\\\hline
\multicolumn{4}{|c|}{redshift: $0.055\leq z<0.1$ / not cluster corrected}\\\hline
 $R$\,[Mpc] & $<\delta_{R,\rm LSB}^2>$ & $<\delta_{R,\rm HSB}^2>$ & $b(R)$\\\hline
  8.0 & 0.612 & 0.745 & 0.906\\ 
  5.6 & 0.643 & 0.835 & 0.877\\
  3.2 & 0.769 & 1.026 & 0.866\\\hline
\multicolumn{4}{|c|}{redshift: $0.01\leq z<0.055$ / cluster corrected}\\\hline
 $R$\,[Mpc] & $<\delta_{R,\rm LSB}^2>$ & $<\delta_{R,\rm HSB}^2>$ & $b(R)$\\\hline
  8.0 & 0.600 & 0.596 & 1.002\\ 
  5.6 & 0.649 & 0.655 & 0.994\\
  3.2 & 0.799 & 0.782 & 1.009\\\hline
\multicolumn{4}{|c|}{redshift: $0.055\leq z<0.1$ / cluster corrected}\\\hline
 $R$\,[Mpc] & $<\delta_{R,\rm LSB}^2>$ & $<\delta_{R,\rm HSB}^2>$ & $b(R)$\\\hline
  8.0 & 0.607 & 0.705 & 0.927\\ 
  5.6 & 0.637 & 0.775 & 0.906\\
  3.2 & 0.756 & 0.931 & 0.902\\\hline\hline
\end{tabular}
\end{center}  
\end{table} 
  
For these particular cases the variance of the density contrast
and the LSB-HSB bias parameter were calculated from environment studies
with spheres of the radius 8.0\,Mpc, 5.6\,Mpc and 3.2\,Mpc.
The first value of $R=8.0$\,Mpc was chosen to
compare the averaged squared density contrast directly with $\sigma_8$ from 
results obtained by other redshift surveys in the literature.
$R=5.6$\,Mpc probes the LSB-HSB bias on scales of the size of large scale
structure filaments ($\sim5$\,Mpc, e.g. \citealt{white87, doroshkevich97}).
The value of $R=3.2$\,Mpc was selected for testing the LSB-HSB clustering
on scales of the diameter of clusters. Furthermore, this radius under-samples 
the averaged size of filaments only marginally. Therefore, it can be  
used to support the results obtained from the bias study of $R=5.6$\,Mpc 
concerning the 
filaments, which of course also shows structure on scales of  $R=3.2$\,Mpc.
It is not reasonable to calculate the bias using studies based on
spheres with smaller radii, because using spheres of a radius $R$ means 
that the local galaxy density is averaged over a sphere of the radius $R$.
If the radius is chosen too small, the fluctuations of the density from galaxy 
to galaxy get too high. Since the bias is a measure of fluctuations,
the study is less meaningful when choosing lower values for $R$. This
can also be explained in another way. 
The radius $R$ of the sphere is a kind of smoothing parameter
for the density field. If one chooses this parameter to be low, the smoothing
effect is too low, and the noise overwhelms the signal.
This effect can directly be seen in Table \ref{bias}. For all cases and
redshift ranges as well as for both galaxy types the variance of the density 
contrast increases with decreasing sphere radii. 

For the study containing field and cluster galaxies within the redshift range
of $0.055\leq z<0.1$ the density contrast for LSBs 
(equation \ref{eq_density_LSBs}) stays clearly below the value obtained for
HSBs (obtained from equation \ref{eq_density_LSBs}) for both the cluster 
corrected case containing only field galaxies as well as the case containing 
cluster and field galaxies. This holds for all
tested scales ($r=8.0, 5.6, 3.2$\,Mpc). Thereby the density contrast 
for LSBs is below that value of HSBs indicating that LSBs are less strongly
clustered than HSBs. This is also seen in the bias parameter $b$. 
For this redshift range it contains values between $b=0.906$ and $b=0.866$
for the cluster and field galaxy case and values in the interval of $b=0.927$ 
and $b=0.902$ for the case containing only field galaxies. These two cases
do not differ a lot. This shows that the galaxy environment of LSBs within the 
redshift interval of $0.055\leq z<0.1$ is clearly less dense than that  
of HSBs for both the cluster corrected and not corrected case.

The first case shows a slightly smaller bias parameter than the 
case containing pure field galaxies. This is caused by the fact that clusters of 
course rise the variance of the density contrast. The sample LSBs 
are not often located in clusters  
(only one LSB of that redshift range was found in a cluster).   
This would explain the increased bias parameter  in the case
of pure field galaxies with respect to the case containing field and cluster
galaxies.
Nevertheless, the bias parameter holds below one for the comparison of
the density contrast between field LSBs and field HSBs within that higher
redshift interval. Taking into account the fact that the LSB population
of that redshift range is dominated by larger LSBs,
this gives strong evidence for a scenario in which the larger type LSBs 
formed and evolved in a lower density region than HSBs.  
Since there is still a lower density contrast for LSBs against that of HSBs in 
the cluster corrected case containing pure field galaxies, this gives strong
support for the initial 
impression from the pie slice (Section \ref{section_pie}). 
This impression was, that for the higher redshift interval $0.05 < z < 0.1$ 
the 
(larger) LSB galaxies are located at the outer rims of the LSS, and some of 
them are even found in void regions. 

The examinations on the density contrast and bias parameter of LSB and HSB
galaxies within the redshift range of $0.01\leq z<0.055$ delivered similar
results except for the cluster corrected case. There, a difference in the
density contrast between these two galaxy populations is not found. 
This results in a bias parameter of $b\sim1$. Taking into account that the LSB 
population in this redshift range is dominated by small, dwarfish LSB galaxies,
one can conclude that these galaxies are found in an environment which
is more similar to that of HSBs. 
However, the average neighbor diagram of that case shows the presence of
a  density contrast (Figure \ref{mean_neighbors_cleaned}, left).

In the case of cluster and
field galaxies (and the same redshift interval of $0.01\leq z<0.055$), a 
significant difference in the density contrast with differences of $\sim0.2$ 
up to  $\sim0.4$ are found. This results in a bias parameter of $b\sim0.86$
indicating that these mainly smaller LSB galaxies do not have the same
clustering properties like cluster and field HSBs combined. 

All in all the galaxy bias study confirms the impression
from the pie slice as well as the results of the mean number of neighbor
diagrams (Figures \ref{mean_neighbors_uncleaned}, 
\ref{mean_neighbors_cleaned}).
From these studies follows that the vicinity of LSB galaxies indeed shows a 
lower galaxy number density than that of HSBs. 
This holds for the sample containing preferably 
larger LSBs as well as the small LSBs. We call this effect the
``LSB-HSB Antibias''.


\section{Discussion \& Conclusions}
As shown above, LSB galaxies reside in a large scale environment with
a lower galaxy density than in the vicinity of HSBs.
This was proved several ways. A first impression to
that fact comes from the pie slice diagram \ref{pie_slice}. 
There, the impression aroused that LSBs in a redshift interval
of $0.05<z<0.1$ are located at the outer parts of the 
filaments and walls. Some LSBs are even found in void regions
of the LSS. The statistical analyses confirm this impression.
For both redshift intervals ($0.01<z<0.055$ and $0.055<z<0.1$) it holds
that LSBs possess less neighbors than HSBs (Figs. 
\ref{mean_neighbors_uncleaned}, \ref{mean_neighbors_uncleaned}, 
sections \ref{environment}, \ref{antibias}). 
This is actually true, if one excludes cluster galaxies and only 
compares the environment of field LSBs to that of field HSBs. 
Even then a stronger isolation of LSBs is found.

First hints to that result already existed in the literature. 
Early results on the studies presented above were 
published in \cite{rosenbaum04}. There it was shown that the 
LSBs found in the Early Data Release (EDR, \citealt{stoughton02} of the 
SDSS possess a lower surrounding galaxy density than HSBs on scales
from 2 to 5\,Mpc.  
Other authors probed the environment of LSBs on 
intermediate and small scales.
It was found that on scales below 2\,Mpc, the galaxy 
environment of LSBs is less dense than that of HSBs (\citealt{bothun93, mo94}).
Furthermore, a lack of nearby  ($r \leq 0.5$\,Mpc) 
companions of LSB galaxies was detected by \cite{zaritsky93}.
However, a study on the distribution of LSBs in the large-scale structure
on scales which correspond to the size of its substructures
(galaxy cluster radii $r\sim$1-3\,Mpc, size of the LSS filaments
and walls $\sim 5$\,Mpc) has not been performed before.

All these results fit well into a formation scenario for LSBs proposed
in \cite{rosenbaum04} which is based on an idea of \cite{bothun97}.
Galaxy formation takes place due to an initial Gaussian spectrum 
of density perturbations with much more low-density fluctuations than high 
density ones.
Many of these low-density perturbations are lost because of the assimilation
or disruption during the evolutionary process of galaxy formation but a 
substantial percentage of the fluctuations survives and is expected to form 
LSB galaxies.
Further on one can assume that the spatial distribution of the initial 
density contrast consists of small scale fluctuations superimposed on
large-scale peaks and valleys.  
Small-scale peaks lead to galaxy formation, whereas the large-scale maxima 
induce cluster and wall formation of the LSS.

If now galaxies formed in the large-scale valleys would develop to LSB 
galaxies because of their isolated environments whereas HSB galaxies 
would form mainly on the large-scale peaks this would lead to 
an universe with LSBs being more isolated than HSBs.

The isolation of LSB galaxies on intermediate and small scales must have 
effected their evolution since tidal encounters acting as triggers of star 
formation would have been rarer in these LSB galaxies than for HSB galaxies. 
Our results give strong evidence for this scenario, since the observed 
isolation of LSB galaxies takes place on scales of $\sim5$\,Mpc, which is 
the typical size of LSS filaments (e.g. \citealt{white87}, 
\citealt{doroshkevich97}). 
Hence,  
LSB galaxies must have formed in the void regions of the LSS.
After that, most 
of them have migrated to the edges of the filaments because of gravitational 
infall, but some of them still remain in the voids where they have formed in.

Another possible explanation for the presence of LSBs 
might be found in the differences in the 
spin parameter $\lambda$ of the Dark Matter halos between LSB and HSB 
galaxies. 
\cite{dalcanton97} studied the formation of disk galaxies and
used a gravitationally self-consistent model for the disk collapse
in order to calculate the observable properties of disk galaxies as a function
of mass and angular momentum of the initial protogalaxy. The observational 
properties of both normal galaxies and LSBs were reproduced. 
Their model generated smooth, asymptotically flat rotation curves
and exponential surface brightness profiles. They found the high angular
momentum halos, which also tended to be low mass, to form naturally low 
baryonic surface density disks or in other words low surface brightness disks.
\cite{boissier03} found in their models for the chemical and 
spectrophotometric evolution of mainly spiral galaxies and LSBs, that 
the models with a spin parameter of $\lambda\sim0.04$ corresponded
to spirals with Freeman values of surface brightness. Moreover, models 
with $\lambda >0.06$ belonged to LSB galaxies. 
From these simulations follows that 
the Dark Matter halos of LSBs possess a higher spin parameter, than that
of HSBs. Since, the spin parameter $\lambda$ is linked to the 
structural properties of the disk,
this would imply a larger scale length for the LSB disks with
respect to HSB disks. Hence, the total gas mass in LSBs would be 
distributed over a larger scale length than in HSBs. This would explain 
the lower gas surface density in LSBs, too.

However, these models for LSB formation do not necessarily contradict
our formation scenario proposed above. 
Moreover, with the latest results of \cite{bailin_05}
it is possible to build a causal connection between this scenario described 
above and the results of simulations,
 that the larger extents of LSB Dark Matter 
halos result in a higher spin parameter.
For these simulations it is assumed that the specific angular
momentum of the baryons is conserved during their dissipation into a 
rotating disk (which is a reasonable assumption). Hence, the scale length of 
the disk would be related to the angular momentum of the Dark Matter halo. 
Therefore, a link would exist between the increased scale length
and the concentration of their Dark Matter halos.
\cite{bailin_05} now performed a cosmological N-body simulation 
with $N=512^3$ on a periodic
50$h^{-1}$\,Mpc volume using the GADGET2 code (by \citealt{springel05b}).
Thereby, $\Lambda$CDM cosmology was assumed. 
The halos of a mass range of $10^{11}\leq M_{vir}/h^{-1}\leq2\cdot10^{12}$
which are typical to be hosts of LSB and HSB galaxies were identified
using a friends-of-friends algorithm and the spin parameter
$\lambda'=J/[\sqrt{2}MVR]$ (with M, the total mass, 
J the total angular momentum, V 
the circular velocity at radius R) in a definition from \cite{bullock01} was 
calculated. As expected, a trend for $\lambda'$ to increase with decreasing
concentration index $c_{200}$ was found. For a concentration index of 
$c_{200}=10$, a median of the spin parameter $\lambda'_{med}\simeq0.03$ and 
for $c_{200}=5$ a value of $\lambda'_{med}\simeq0.05$ was obtained 
by the authors.  Then, assuming angular momentum
conservation of the baryons during dissipation into the rotating disk,
the disk angular momentum and its extent for a given mass was calculated.
From that, the central surface densities were estimated and
the central surface brightnesses were estimated by using the
fitting equations of \cite{mo98}. At the end they found a clear trend 
that halos with lower concentrations host disks with
lower central surface densities. With these results, for the first time a 
correlation between the spin parameter $\lambda'$ and the concentration of 
the Dark Matter halos was found.
The obtained results are 
not only important for the theories which predict the existence of LSBs
to be caused by the higher spin parameter of Dark Matter halos. They are
also important for the results of section \ref{environment}, since
this could be the connection between these two scenarios.
The type of diffuse, less concentrated Dark Matter halos, which preferably 
host LSB galaxies, one would expect to exist in lower density regions,
whereas the more concentrated Dark Matter halos one would expect to
be formed in high density regions of the initial universe.
This is compatible to the results of \cite{avila_reese05}, who
performed $\Lambda$CDM  
N-body simulations. They found the halos in clusters to have a lower median
spin parameter, to be more spherical, 
and to possess less aligned internal angular momentum 
than the halos in void or field regions. Their simulations
showed trends that disk galaxies 
which formed in halos with low spin parameters, 
but high concentration indices, are preferably of earlier morphological types.
Furthermore, these galaxies have higher surface brightnesses, 
smaller scale lengths, and lower gas fractions than galaxies formed in halos 
which have higher spin parameters, but are low-concentrated.

Hence, the results of \cite{bailin_05} in combination with the
simulations of \cite{avila_reese05} and the results presented
in the present publication answer the following question:  Are LSB galaxies 
caused by nature or nurture? They exist due to a mixture of both.

\begin{acknowledgements}
We are grateful to Clemens Trachternach for 
proof-reading and precious incitations to this paper. 
This work was supported financially by the GRK\,787 ``Galaxy Groups as 
Laboratories for Baryonic and Dark Matter'', and DFG project BO1642/4-1.
Funding for the creation and distribution of the SDSS Archive has been 
provided by the Alfred P. Sloan Foundation, the Participating Institutions, 
the NASA, the National Science Foundation, the US Department of Energy, the 
Japanese Monbukagakusho, and the Max Planck Society. 
\end{acknowledgements}






   
  


\bibliographystyle{aa}
\bibliography{07462}


\end{document}